\documentclass[journal,comsoc]{IEEEtran}

\usepackage[T1]{fontenc}

\usepackage{ifpdf}

\usepackage{cite}

\usepackage[pdftex]{graphicx}

\usepackage{amsmath}
\usepackage[cmintegrals]{newtxmath}

\usepackage{bm}

\usepackage{algorithmic}

\usepackage{array}

\ifCLASSOPTIONcompsoc
 \usepackage[caption=false,font=normalsize,labelfont=sf,textfont=sf]{subfig}
\else
 \usepackage[caption=false,font=footnotesize]{subfig}
\fi

\usepackage{url}

\begin{document}

\title{User-generated Video Quality Assessment: A Subjective and Objective Study}

\author{Yang~Li,
		Shengbin~Meng,
		Xinfeng~Zhang,~\IEEEmembership{Member,~IEEE,}
		Shiqi~Wang,~\IEEEmembership{Member,~IEEE,}
		Yue~Wang,
		Siwei~Ma,~\IEEEmembership{Member,~IEEE,}
\thanks{Yang Li and Siwei Ma are with the Institute of Digital Media, Peking University, Haidian District, Beijing 100871, China.}
\thanks{Xinfeng Zhang is with the School of Computer Science and Technology, University of Chinese Academy of Sciences, Beijing 101408, China.}
\thanks{Shiqi Wang is with Department of Computer Science, City University of
Hong Kong, Kwoloon, HK, China.}
\thanks{Shengbin Meng and Yue Wang are with media foundation team, Bytedance Inc.}}

%
%

\markboth{Journal of \LaTeX\ Class Files,~Vol.~14, No.~8, August~2015}%
{Shell \MakeLowercase{\textit{et al.}}: Bare Demo of IEEEtran.cls for IEEE Communications Society Journals}
%


\maketitle

\begin{abstract}
Recently, we have observed an exponential increase of user-generated content (UGC) videos.
The distinguished characteristic of UGC videos originates from the video production and delivery chain, as they are usually acquired and processed by non-professional users before uploading to the hosting platforms for sharing. As such, these videos usually undergo multiple distortion stages that may affect visual quality before ultimately being viewed. Inspired by the increasing consensus that the optimization of the video coding and processing shall be fully driven by the perceptual quality, in this paper, we propose to study the quality of the UGC videos from both objective and subjective perspectives. We first construct a UGC video quality assessment (VQA) database, aiming to provide useful guidance for the UGC video coding and processing in the hosting platform. The database contains source UGC videos uploaded to the platform and their transcoded versions that are ultimately enjoyed by end-users, along with their subjective scores. 
Furthermore, we develop an objective quality assessment algorithm that automatically evaluates the quality of the transcoded videos based on the corrupted reference, which is in accordance with the application scenarios of UGC video sharing in the hosting platforms. The information from the corrupted reference is well leveraged and the quality is predicted based on the inferred quality maps with deep neural networks (DNN). Experimental results show that the proposed method yields superior performance. Both subjective and objective evaluations of the UGC videos also shed lights on the design of perceptual UGC video coding.


\end{abstract}


%
\IEEEpeerreviewmaketitle

\section{Introduction}
\label{sec:introduction}

\IEEEPARstart{V}{ideo} content was historically created by professional content producers. Recently, with the development of multimedia and network technologies, as well as the advances of acquisition devices, there has been an explosion of user-generated content (UGC) videos and related sharing services. Enormous videos generated without professional routines and practices are uploaded to sharing platforms such as Facebook, YouTube and TikTok. 
Comparing to professionally-generated content (PGC) videos, the low barriers in video production and sharing make the UGC content extremely diverse. In particular, the lack of proper shooting skills and professional video capture equipment make the perceptual quality of UGC videos even worse. Besides, special effects are sometimes incorporated to enhance the user experience, thereby increasing the difficulty of quality assessment and compression.
Exponential increase in the demand for high-quality videos poses great challenges in practice. As such, effective UGC video quality assessment (VQA) algorithms become critical to guide the optimization of the hosting platform, in an effort to deliver videos with better visual quality under limited bandwidth.

In the traditional full-reference (FR) quality assessment, pristine sources are available for reference, 
such that the quality of the distorted video can be predicted by signal or feature level comparisons. 
However, straightforwardly applying this strategy to UGC videos is problematic, as the source videos in the hosting platform have already been corrupted due to acquisition and compression distortions introduced before uploading to the hosting platform. 
As such, the traditional FR algorithms may be mislead by the distorted reference and fail to predict the quality of the ultimately viewed UGC videos. 
One extreme example is that an excessively high bit rate is applied to transcode a video with extremely poor quality. 
In this scenario, the objective FR quality is not consistent with the subjective quality due to the high similarity with the corrupted reference.
However, relying on no-reference (NR) algorithms only may omit the useful reference information, and may not be able to ensure the accurate prediction with high robustness on such diverse content.

In this paper, we first creat a database with subjective ratings for UGC videos, revealing the complex nature of the UGC quality assessment problem. Furthermore, we present a corrupted-reference framework which delivers accurate predictions of the perceptual quality for the UGC videos. The proposed algorithm measures perceptual quality by combining the local distortions of the source and transcoded videos relying on the prediction of the quality maps. In particular, the quality maps are predicted in a data-driven manner, and fused through a learned network such that the overall quality score is estimated by gradually pooled features. The three main contributions of this work are as follows.
\begin{enumerate}
  \item We construct a dedicated exploration database for UGC videos, including the source and transcoded videos in the hosting platforms, as well as the objective and subjective scores. We further demonstrate that innovative quality assessment approaches should be developed based on careful investigations. 
  \item We propose a novel corrupted-reference VQA method for UGC videos based on deep neural networks (DNN). In contrast with traditional FR quality models, the intrinsic quality of the corrupted-reference is incorporated to accurately infer the quality.
  \item We show that the performance of the proposed framework outperforms the state-of-the-art methods in the application domain of UGC video processing. While the field of UGC video coding and processing is still quickly evolving, we also envision the future perceptual UGC video compression scheme based on the proposed quality measure. 
\end{enumerate}



\section{Related works}
\label{sec:relatedwork}

\subsection{Objective VQA Measures}
\label{ssec:vqa}


\subsubsection{Full-Reference VQA}
\label{ssec:frvqa}

FR VQA algorithms deliver robust and accurate predictions based on the fully accessible reference information.
In~\cite{seshadrinathan2011temporal}, hysteresis effect in the subjective testings is observed, and a hysteresis based temporal pooling strategy is applied to extend image quality assessment (IQA) metrics such as PSNR and SSIM~\cite{SSIM} to VQA, which has been proved to be better than average pooling.
In~\cite{wang2004video} and~\cite{wang2012novel}, video quality measures have been designed based on structural features.
Lu \emph{et al.}~\cite{lu2019spatiotemporal} described the degradation of video quality via spatiotemporal 3D gradient differencing.
Moreover, a VQA algorithm based on statistical characteristics of optical flows was proposed in~\cite{manasa2016optical}.
In~\cite{seshadrinathan2009motion}, a spatio-spectrally localized multiscale framework for evaluating dynamic video fidelity by motion quality along computed motion trajectories was presented.
In~\cite{VIS3}, ViS3 estimates quality via separate predictions of perceived degradation  originated from spatial distortion as well as joint spatial and temporal distortion.
Machine learning has also played a critical role in the development of modern VQA models.
In~\cite{freitas2018using}, several perceptual-relevant features and methods have been combined by random forest regression algorithm to boost the performance.
In~\cite{VMAF}, video multi-method assessment fusion (VMAF) produces remarkably improved  quality prediction performance by mapping multiple features to human-quality opinions using support vector regressor (SVR).
Motivated by the great success of convolutional neural network (CNN) on numerous visual analysis tasks, a DNN based approach has been developed by joint learning of local quality and local weights in~\cite{bosse2017deep}, and a pairwise-learning framework was proposed in~\cite{prashnani2018pieapp} to train a perceptual image-error metric. 
Kim \emph{et al.}~\cite{kim2018deep} quantified the spatio-temporal visual perception via a CNN and a convolutional neural aggregation network, and Zhang \emph{et al.}~\cite{zhang2019objective} proposed a FR VQA metric by integrating transfer learning with a CNN.

\subsubsection{No-Reference VQA}
\label{ssec:nrvqa}

NR VQA is a more natural and preferable way to assess the perceived video quality as the reference videos are unavailable in many practical video applications.
Many methods focus on estimating the perceived quality of videos with specific distortions, such as compression distortion~\cite{zhu2014no}, transmission error~\cite{zhang2012additive} and scaling artifacts~\cite{ghadiyaram2017no}.
For distortion-unaware NR VQA methods, natural scene statistics (NSS) or natural video statistics (NVS) models are usually used as they are sensitive to diverse distortions.
Saad \emph{et al.}~\cite{saad2014blind} proposed a NR VQA algorithm, known as VBLIINDS, which contains a NSS model and a motion model that quantifies motion coherency.
Mittal \emph{et al.}~\cite{VIIDEO} proposed a VQA model termed as the video intrinsic integrity and distortion evaluation oracle (VIIDEO), which quantifies disturbances introduced due to distortions according to the NVS model.
In \cite{zhu2017blind}, the video content is disassembled into the predicted part and the uncertain part, such that their quality degradations are separately evaluated by NVS model to yield the overall quality.
Li \emph{et al.}~\cite{li2016spatiotemporal} proposed a NR-VQA metric based on NVS in the 3D discrete cosine transform (3D-DCT) domain. Recently, CNN based NR-VQA methods have also been developed. 
Li \emph{et al.}~\cite{li2015no} proposed a shearlet- and CNN-based NR VQA (SACONVA), where spatiotemporal features extracted by 3D shearlet transform are fed to a CNN to predict a perceptual quality score.
Liu \emph{et al.}~\cite{liu2018end} exploited the 3D-CNN model for codec classification and quality assessment of compressed videos.
In~\cite{zhang2018blind}, a NR VQA framework based on weakly supervised learning with a CNN and a resampling strategy was presented.
Li \emph{et al.}~\cite{li2019quality} proposed a NR-framework for in-the-wild videos by incorporating content dependency and temporal-memory effects.
Moreover, generative networks have also been used to predict the quality map or source given the distorted image to help the blind IQA task~\cite{ren2018ran4iqa, pan2018blind}. 

\subsection{VQA Databases}
\label{ssec:vqadatabase}

There are several publicly available video databases for VQA. 
LIVE~\cite{LIVE} collects 10 uncompressed high-quality videos as reference videos, and correspondingly 150 distorted videos were created using four different distortion types and  strengths. 
LIVE Mobile~\cite{LIVE-mobile} consists of 200 distorted videos created from 10 RAW HD reference videos, and dynamically varying distortions are also considered.
In MCL-JCV~\cite{MCL-JCV}, a compressed VQA database was created based on the just noticeable difference (JND) model.
CVD2014~\cite{CVD2014} contains a total of 234 videos that are recorded using 78 different cameras, along with open-ended quality descriptions such as sharpness, graininess and color balance provided by the observers.
LIVE-Qualcomm~\cite{LIVE-incapture} consists of 208 videos captured using 8 different mobile devices which model six common in-capture distortion categories.
KoNViD-1k~\cite{konvid} is a subjectively VQA database containing 1,200 public-domain videos that are fairly sampled from a large public video database, YFCC100M.
LIVE-VQC~\cite{LIVE-VQC} contains 585 videos captured using 101 different devices with a wide range of distortion levels.

Apparently, databases with high quality source videos such as LIVE and LIVE Mobile may not align with the UGC application scenarios, where databases with acquisition distortion  such as CVD2014, LIVE-Qualcomm and KoNViD-1k are more realistic. 
In~\cite{yu2019predicting}, LIVE Wild Compressed Picture Quality Database has been constructed, where images with acquisition distortions are further compressed. 
However, database dedicated to UGC videos by considering the UGC video compression still remains absent and there is a strong desire for an adequate database sufficing to simulate the UGC production chain from acquisition to processing on the hosting platform. 

\section{UGC Video Database}
\label{sec:database}


\subsection{Video Collection}
\label{ssec:videocollection}

To cover typical content and characteristics representing UGC videos, 400 videos are randomly selected from the videos uploaded to TikTok~\cite{tiktok} that meet the following criteria:
\begin{itemize}
\item With a resolution of 1280$\times$720 (height$\times$width);
\item Belonging to the category of selfie, indoor, outdoor or screen content.
\item Last longer than 10 seconds;
\item Played at 30 frames per second (FPS);
\end{itemize}

Since 720p is one of the most widely adopted UGC video formats, we ensure that all selected videos share this resolution. 
Most videos can be classified into one of the selfie, indoor, outdoor and screen content videos. In particular, most areas of selfie videos are occupied by human face, and screen content videos are mainly game screen recording. Moreover, indoor videos are life scenes shot in close-up, and outdoor videos are outdoor scenery acquired with a distant view. A few videos with special content have been filtered out. Since we will crop all these videos to 10 seconds, videos shorter than 10 seconds are not considered here.

\subsection{Video Sampling}
\label{ssec:videosampling}

\begin{figure}[t]
\centering
\subfloat[]{\includegraphics[width=0.90\columnwidth]{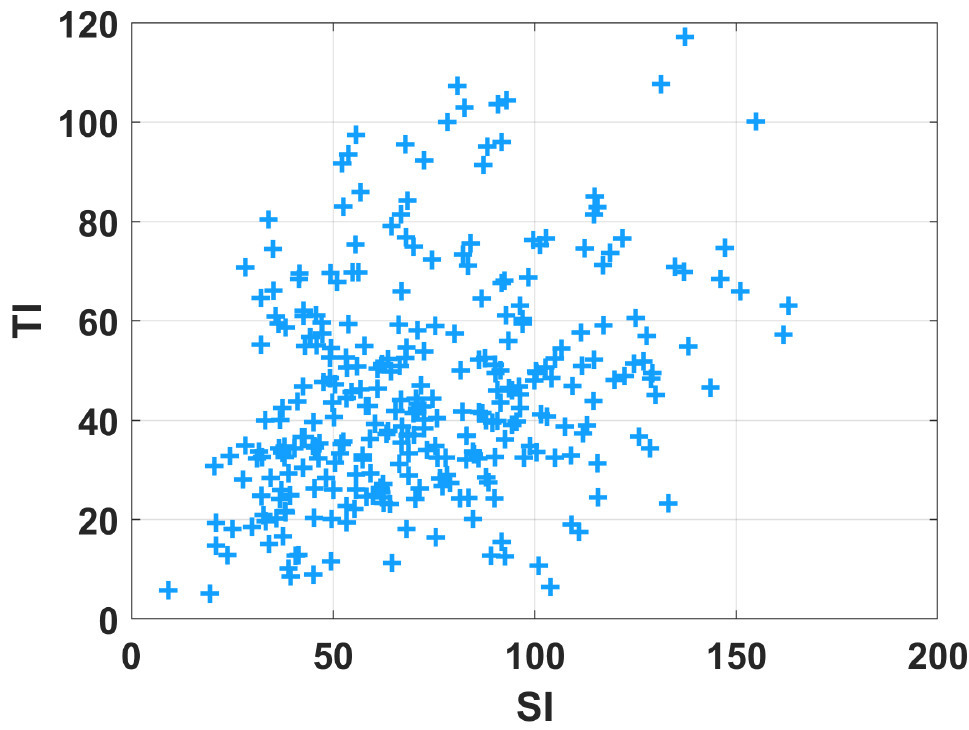}%
\label{fig:siti400}}
\hfil
\subfloat[]{\includegraphics[width=0.90\columnwidth]{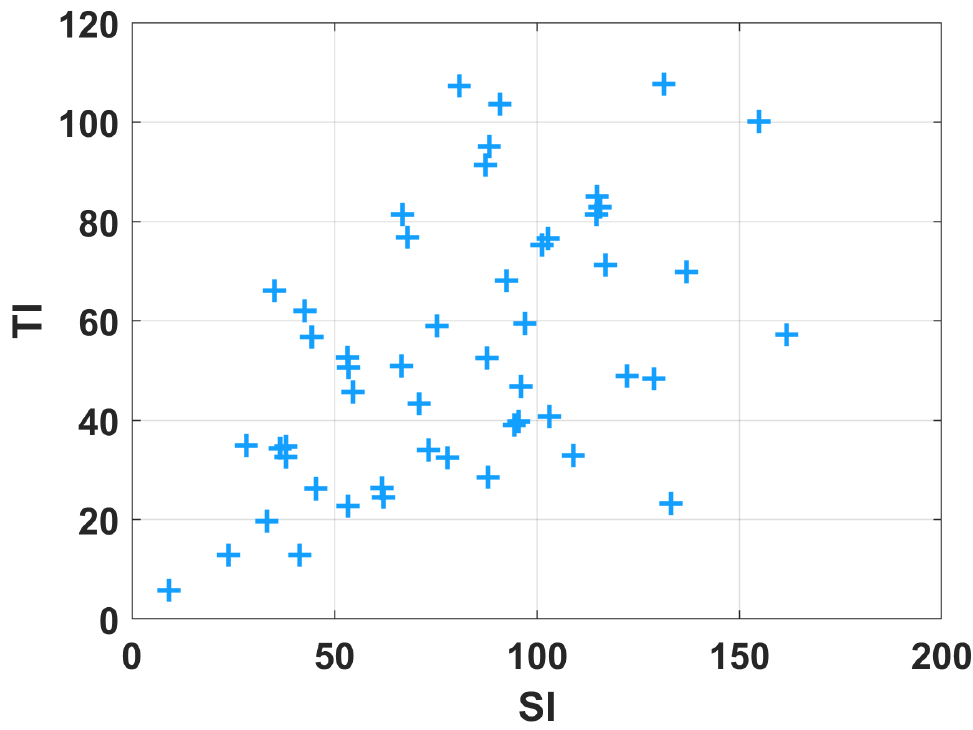}%
\label{fig:siti50}}
\caption{Distribution of SI and TI indices for UGC videos. \protect\subref{fig:siti400} 400 videos before sampling; \protect\subref{fig:siti50} 50 videos after sampling.}
\label{Fig:SITI}
\end{figure}

Subsequently, we sample the videos selected in the previous step according to the statistical characteristics of videos to obtain the final source videos. Specifically, three attributes including spatial perceptual information (SI), temporal perceptual information (TI) and blur index have been employed. Among these indicators, SI and TI are highly correlated with the levels of distortion when the video is lossy transmitted, as suggested in \cite{rec2008p}. Since UGC videos uploaded by users are usually accompanied by varying degrees of blurry artifacts which significantly affect the perceptual quality, the blur metric is also included.

\emph{SI:} SI quantifies spatial complexity and variety of a video, and it is defined as the maximum standard deviation over all Sobel-filtered frames, 
\begin{equation}
\label{eq:si}
SI=max_{time}{\{std_{space}[Sobel(F_n)]\}}
\end{equation}
where $F_n$ represents frame $n$, $Sobel(\cdot)$ is Sobel filter and $std_{space}$ represents the standard deviation over space. 

\emph{TI:} TI quantities the temporal changes of a video, and it is given by the maximum standard deviation of the frame difference derived from adjacent frames. As such, it can be formulated as follows,
\begin{equation}
\label{eq:ti}
TI=max_{time}{\{std_{space}[M_n(i,j)]\}}
\end{equation}
\begin{equation}
\label{eq:mnij}
M_n(i,j)=F_n(i,j)-F_{n-1}(i,j)
\end{equation}
where $F_n(i,j)$ is pixel value at $(i,j)$ of frame $n$.

\emph{Blur:} The cumulative probability of blur detection (CPBD) indicator~\cite{narvekar2011no} is adopted here to evaluate the levels of blur. The average CPBD value of the sampled frames is used to indicate the blurriness of the video.

\begin{figure*}[t]
\centering
\subfloat[]{\includegraphics[width=0.12\textwidth]{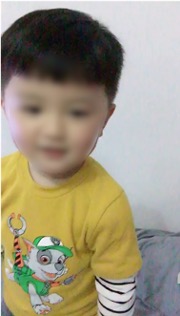}%
\label{fig:selfie1}}
\hfil
\subfloat[]{\includegraphics[width=0.12\textwidth]{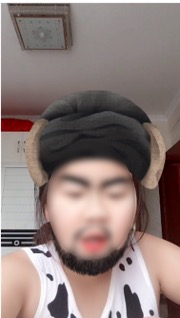}%
\label{fig:selfie2}}
\hfil
\subfloat[]{\includegraphics[width=0.12\textwidth]{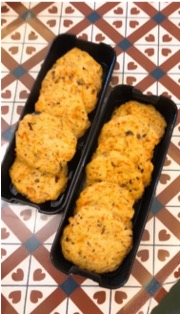}%
\label{fig:indoor1}}
\hfil
\subfloat[]{\includegraphics[width=0.12\textwidth]{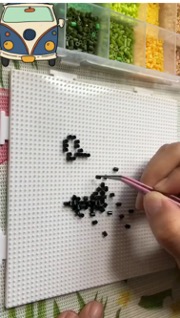}%
\label{fig:indoor2}}
\hfil
\subfloat[]{\includegraphics[width=0.12\textwidth]{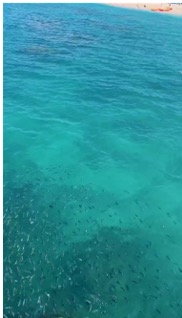}%
\label{fig:outdoor1}}
\hfil
\subfloat[]{\includegraphics[width=0.12\textwidth]{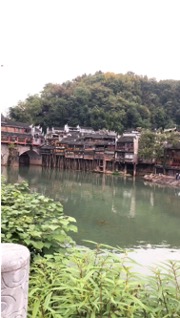}%
\label{fig:outdoor2}}
\hfil
\subfloat[]{\includegraphics[width=0.12\textwidth]{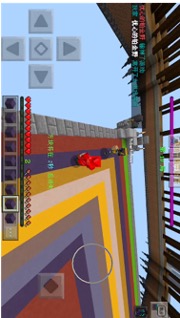}%
\label{fig:screen1}}
\hfil
\subfloat[]{\includegraphics[width=0.12\textwidth]{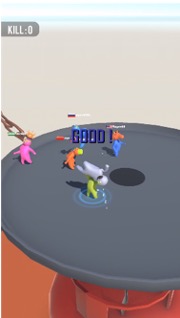}%
\label{fig:screen2}}
\caption{Examples of videos in the database. \protect\subref{fig:selfie1}-\protect\subref{fig:selfie2}: selfie videos; \protect\subref{fig:indoor1}-\protect\subref{fig:indoor2}: indoor videos; \protect\subref{fig:outdoor1}-\protect\subref{fig:outdoor2}: outdoor videos; \protect\subref{fig:screen1}-\protect\subref{fig:screen2}: screen content videos.}
\label{Fig:database}
\end{figure*}

Before sampling from these videos, we crop these videos to 10 seconds and remove the audio parts. To enable the characteristics of sampled videos uniformly distributed in terms of these features, we adopt the sampling strategy introduced in~\cite{vonikakis2016shaping}.
In particular, the original videos are characterized with a set \(S\),
\begin{equation}
\label{eq:origdata}
    S=\left\{ \bm{q}_i|\bm{q}_i\in \mathbb{R}^M,\bm{q}_i\sim D_S^M \right\}_{i=1}^K
\end{equation}
where $M$ and $K$ represent the number of features and videos, respectively (here $M=3, K=400$).
The main objective is to select a subset of $N$ videos,
\begin{equation}
    s=\left\{ \bm{\hat q}_i|\bm{\hat q}_i\in S,\bm{\hat q}_i\sim D_s^M \right\}_{i=1}^N
\end{equation}
with the uniform distribution $\bm D\in \mathbb{R}^{H\times M}$ (each of its columns $\bm D_{*j}$ denoting the probability mass function across the $j^{th}$ dimension which is quantized into \(H\) bins).
As such, we introduce a set of $M$ binary matrices $B=\left\{\bm{B}^m \right\}_{m=1}^M$, in which $b_{ij}^m$ denotes whether or not the $j^{th}$ item of $S$ belongs to $i^{th}$ interval of the target PMF for the dimension \(m\), and binary vector $\bm x\in \mathbb{Z}_2^K$, where $x_i$ is decision variable determining whether $i^{th}$ item of $S$ belongs to subset $s$.
As such, this problem can be formulated as follows,
\begin{equation}
\label{eq:opti}
  \mathop{\min}_{\bm{x}} \sum_{m=1}^M\| \bm{B}^m \bm x - N\bm D_{*m}\|_1 \ \ s.t.\ \|\bm x\|_1=N .
\end{equation}
By finding the best solution with the optimization objective, a subset that is closest to the uniform distribution on all features can be sampled from the original database. Finally, 12, 13, 13, 12 videos were chosen from selfie, indoor, outdoor and screen content videos, respectively, using this sampling strategy with $H=5$. 
Fig.~\ref{Fig:SITI} shows the plots of SI against TI for 400 videos before sampling and 50 videos after sampling, which is apparent that sampled videos  span a wide range of SI-TI spaces. Moreover, the snapshots of some example sampled videos from each content category are shown in Fig.~\ref{Fig:database}.

\subsection{Video Transcoding}
\label{ssec:distortedvideo}

Considering that our primary goal of building this database is to simulate the UGC production chain from acquisition to processing on the hosting platform, based on which quality assessment algorithm can be developed in an effort to further improve the transcoding performance, we further transcode these sampled source videos using different codecs and compression levels. More specifically, H.264/AVC~\cite{h264} encoder x264~\cite{x264} and HEVC~\cite{h265} encoder x265~\cite{x265} have been used to simulate the transcoding process in the hosting platforms, and five common used quantization parameters (QPs) 22, 27, 32, 37, 42 are used to control the quality level of transcoded videos for each codec. As such, each source video can be transcoded to 10 corresponding transcode versions. Finally, there are 550 videos in our built UGC-VIDEO database including source videos.

\begin{table}[t]
\centering
\caption{Performance comparisons of quality assessment algorithms in terms of SROCC.}
\label{tab:benchmark_srocc}
\begin{tabular}{|c|c|c|c|c|c|}
\hline
Methods   & Selfie  & Indoor  & Ourdoor & Screen  & Full database \\ \hline
BRISQUE   & 0.436   & 0.327   & 0.580   & 0.346   & 0.354         \\ \hline
NIQE      & 0.511   & 0.480   & 0.453   & 0.128   & 0.314         \\ \hline
VIIDEO    & 0.113   & 0.348   & 0.218   & 0.026   & 0.085         \\ \hline
BLIINDS   & 0.382   & 0.386   & 0.051   & 0.462   & 0.175         \\ \hline
PSNR      & 0.715   & 0.700   & 0.664   & 0.489   & 0.612         \\ \hline
VIF       & 0.837   & 0.803   & 0.807   & 0.629   & 0.736         \\ \hline
SSIM      & 0.842   & 0.798   & 0.857   & 0.464   & 0.714         \\ \hline
MS-SSIM   & 0.821   & 0.783   & 0.842   & 0.507   & 0.722         \\ \hline
SpEED-QA  & 0.839   & 0.747   & 0.838   & 0.746   & 0.786         \\ \hline
ViS3      & 0.762   & 0.706   & 0.823   & 0.699   & 0.746         \\ \hline
VMAF      & 0.823   & 0.821   & 0.856   & 0.825   & 0.814         \\ \hline
\end{tabular}
\end{table}

\begin{table}[t]
\centering
\caption{Performance comparisons of quality assessment algorithms in terms of PLCC.}
\label{tab:benchmark_plcc}
\begin{tabular}{|c|c|c|c|c|c|}
\hline
Methods   & Selfie  & Indoor  & Ourdoor & Screen  & Full database \\ \hline
BRISQUE   & 0.416   & 0.346   & 0.611   & 0.328   & 0.315   \\ \hline
NIQE      & 0.509   & 0.511   & 0.520   & 0.056   & 0.176   \\ \hline
VIIDEO    & 0.251   & 0.178   & 0.326   & 0.032   & 0.157   \\ \hline
BLIINDS   & 0.415   & 0.421   & 0.001   & 0.464   & 0.216   \\ \hline
PSNR      & 0.717   & 0.733   & 0.639   & 0.452   & 0.579   \\ \hline
VIF       & 0.862   & 0.820   & 0.850   & 0.633   & 0.626   \\ \hline
SSIM      & 0.866   & 0.847   & 0.857   & 0.590   & 0.769   \\ \hline
MS-SSIM   & 0.845   & 0.841   & 0.865   & 0.626   & 0.773   \\ \hline
SpEED-QA  & 0.748   & 0.671   & 0.730   & 0.724   & 0.673   \\ \hline
ViS3      & 0.787   & 0.744   & 0.872   & 0.754   & 0.783   \\ \hline
VMAF      & 0.884   & 0.886   & 0.907   & 0.830   & 0.863   \\ \hline
\end{tabular}
\end{table}

\subsection{Subjective Testing and Analyses}
\label{ssec:subtesting}

After collecting the videos, subjective testing is further conducted to obtain the subjective scores using 
absolute category rating with hidden reference (ACR-HR)~\cite{itu1999subjective} method, in which the videos are played one by one and the subjects are asked to provide a opinion score according the five-grade rating scales. 
The full database is divided into three sessions, each containing 16 or 17 source videos along with their respective transcoded versions. Hence, each session lasts about half an hour to minimize viewer fatigue. In particular, at the beginning of each session, ``dummy presentations'' with various levels of perceptual quality have been introduced to stabilize the opinion of subjects and the opinion data of these presentations are not taken into account in the final result of the experiment.
The videos are displayed at their original resolution without scaling, and the subjects are required to click the corresponding button within a few seconds to choose from ``Excellent'',``Good'', ``Fair'', ``Poor'' and ``Bad'', corresponding to 5\(\sim\)1 points. 
A total of 28 subjects participated in this test. Since this is a hidden-reference study, the source videos have also been included in the subjective testings. As such, besides the mean opinion score (MOS), the differential mean opinion score (DMOS) can also be obtained. 

\begin{figure*}[ht]
\centering
\subfloat[MOS: 4.43]{\includegraphics[width=0.24\textwidth]{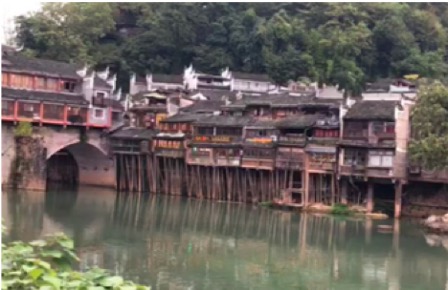}%
\label{fig:fr_limit_a}}
\hfil
\subfloat[MOS: 4.11]{\includegraphics[width=0.24\textwidth]{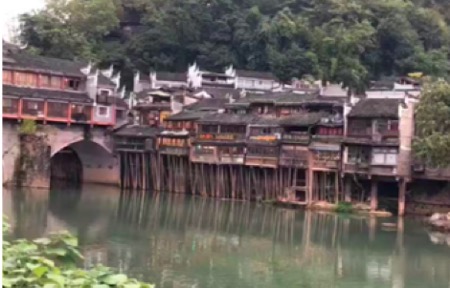}%
\label{fig:fr_limit_b}}
\hfil
\subfloat[MOS: 3.21]{\includegraphics[width=0.24\textwidth]{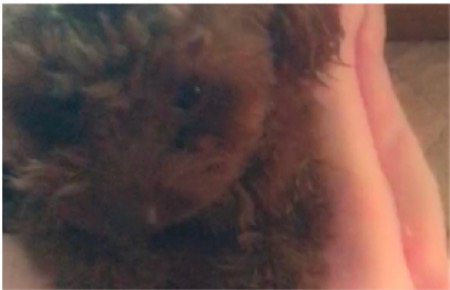}%
\label{fig:fr_limit_c}}
\hfil
\subfloat[MOS: 2.96]{\includegraphics[width=0.24\textwidth]{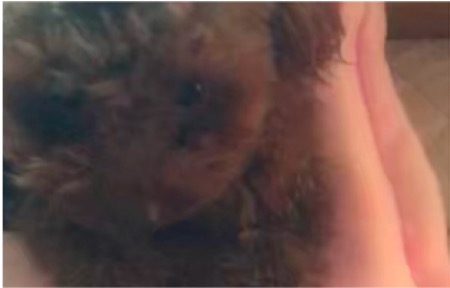}%
\label{fig:fr_limit_d}}
\caption{Performance of FR metrics on the proposed database. \protect\subref{fig:fr_limit_a} High quality reference video; \protect\subref{fig:fr_limit_b} HEVC transcoded video of \protect\subref{fig:fr_limit_a}, PSNR: 41.64dB, SSIM: 0.996; \protect\subref{fig:fr_limit_c} Low quality reference video; \protect\subref{fig:fr_limit_d} HEVC transcoded video of \protect\subref{fig:fr_limit_c}, PSNR: 41.73dB, SSIM: 0.977.}
\label{Fig:fr_limitation}
\end{figure*}

\begin{figure*}[t]
\centering
\includegraphics[width=0.95\textwidth]{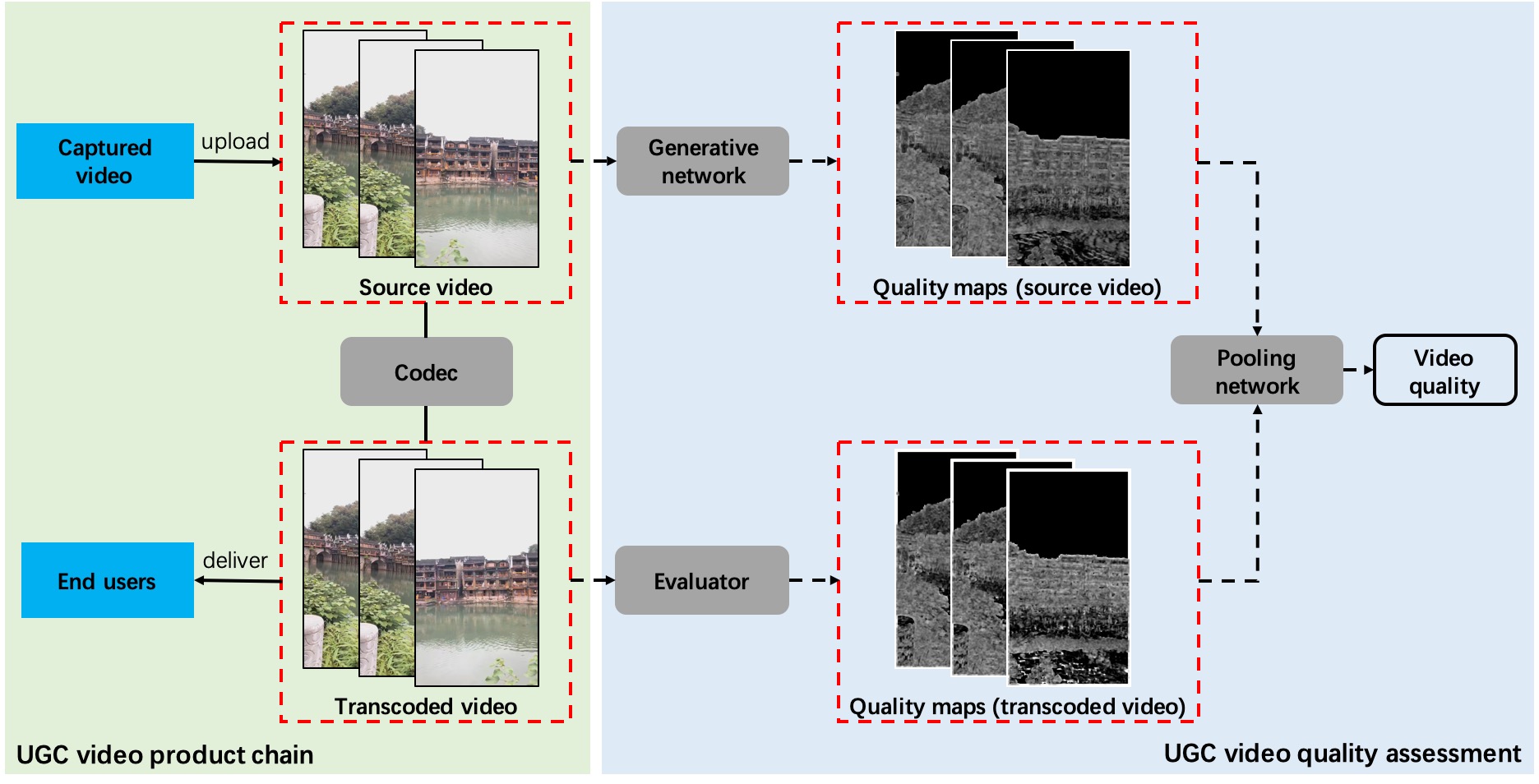}
\caption{Framework of the proposed objective quality assessment method. The quality maps from the source videos as well as the comparisons between source and transcoded videos are fused with a pooling network to obtain the final quality.}
\label{Fig:framework}
\end{figure*}


The screening of subjects is further conducted as specified in ITU-R BT 500.13~\cite{series2012methodology}.
The kurtosis of scores are calculated to determine if the scores for each test presentation are normally distributed. Score range of each video is then computed as 2 or $\sqrt{20}$ standard deviations from the mean scores according to whether the scores are normally distributed. For each subject $i$, we count the number of scores above and blow this range, denoted as $P_i$ and $Q_i$.
As such, the subject $i$ will be rejected when,
\begin{equation}
  \frac {P_i+Q_i}{JK}>0.05 \ \ \text{and} \ \ |\frac{P_i-Q_i}{P_i+Q_i}|<0.3
\end{equation}
where $J$ is the number of versions for each source and $K$ denotes the number of source videos. 
Based on the our analysis, no subject has been rejected at this stage.

\subsection{Performance of Existing Models}
\label{ssec:benchmark}

We evaluate the performance of several objective quality assessment algorithms on the established database using Spearman's rank ordered correlation coefficient (SROCC) and Pearson's linear correlation coefficient (PLCC).
In particular, the larger the values of SROCC and PLCC, the better the performance.
Besides, before computing PLCC, the predicted scores are passed through a logistic non-linearity regression as suggested in ~\cite{sheikh2006statistical}:
\begin{equation}\label{eqn:fitting}
f(x) = \beta_1(\frac{1}{2}-\frac{1}{1+e^{\beta_2(x-\beta_3)}})+\beta_4x+\beta_5.
\end{equation}
IQA algorithms are extended to VQA methods by averaging frame-level quality scores. The tested quality measures include PSNR, SSIM~\cite{SSIM}, MS-SSIM~\cite{MS-SSIM}, VIF~\cite{VIF}, SpEED-QA~\cite{bampis2017speed}, ViS3~\cite{VIS3}, VMAF~\cite{VMAF}, VBLIINDS~\cite{saad2014blind} and VIIDEO~\cite{VIIDEO}.
Table~\ref{tab:benchmark_srocc} and Table~\ref{tab:benchmark_plcc} tabulate the SROCC and PLCC between the algorithm scores and MOS for each content category, as well as across the full database. It is disappointing to find that the existing algorithms may not be able to provide adequate predictions on the UGC videos. However, these results still provide some useful insights that could benefit the design of the UGC VQA models. First, as illustrated in Fig. 3, the quality based on comparisons against the reference could be problematic due to the corrupted reference. This suggests the importance of including the intrinsic quality of the reference. Second, most of the tested algorithms perform the worst on screen content videos, and this may be attributed to the particularity of this type of videos compared to natural videos. 
These observations motivate a specifically designed VQA model that equips the intrinsic quality of the corrupted reference as well as the data-driven model for learning the statistics of the video content.


\begin{figure*}[ht]
\centering
\subfloat[]{\includegraphics[width=0.12\textwidth]{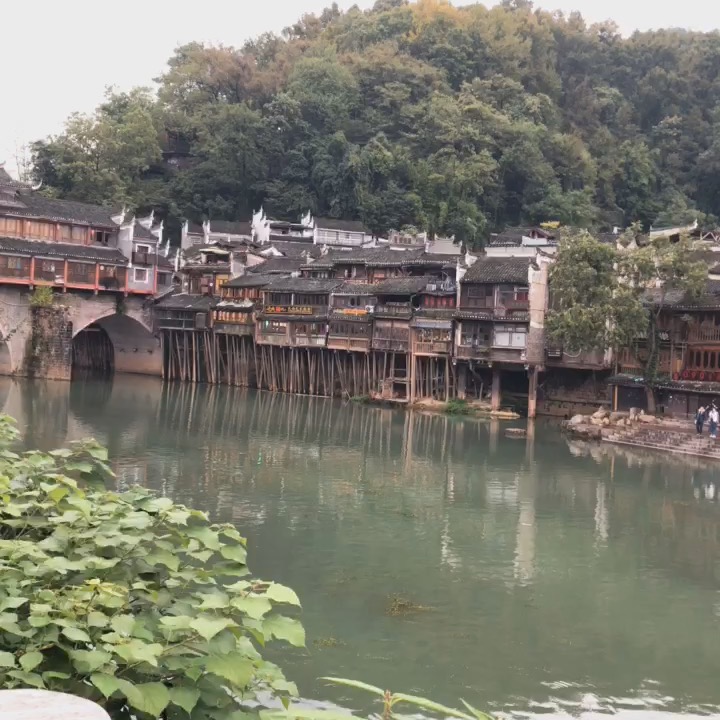}%
\label{fig:fr_map_ref}}
\\
\subfloat[]{\includegraphics[width=0.12\textwidth]{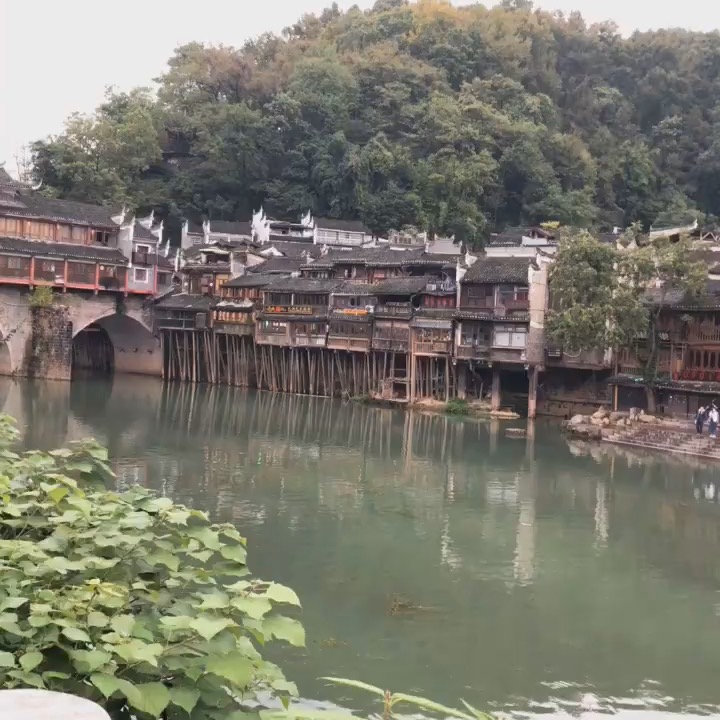}%
\label{fig:fr_map_dist_1}}
\hfil
\subfloat[]{\includegraphics[width=0.12\textwidth]{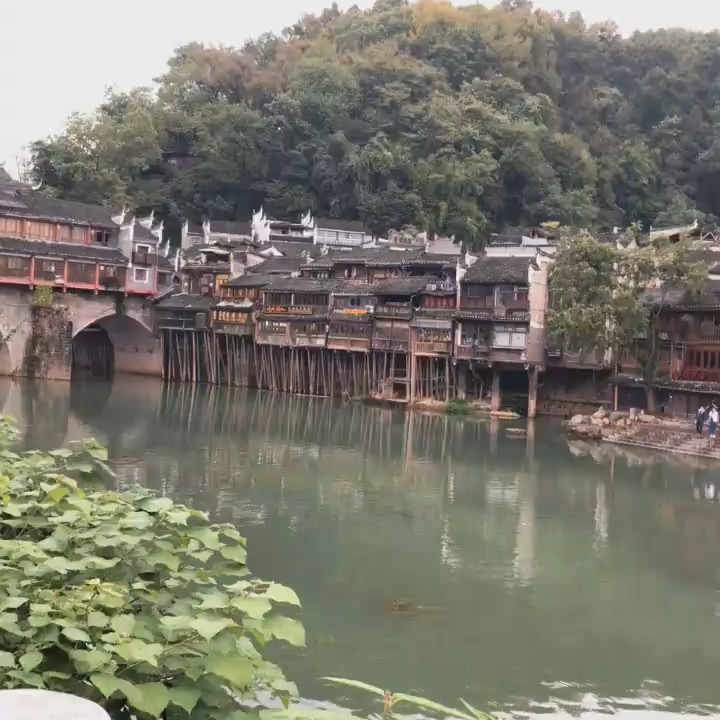}%
\label{fig:fr_map_dist_2}}
\hfil
\subfloat[]{\includegraphics[width=0.12\textwidth]{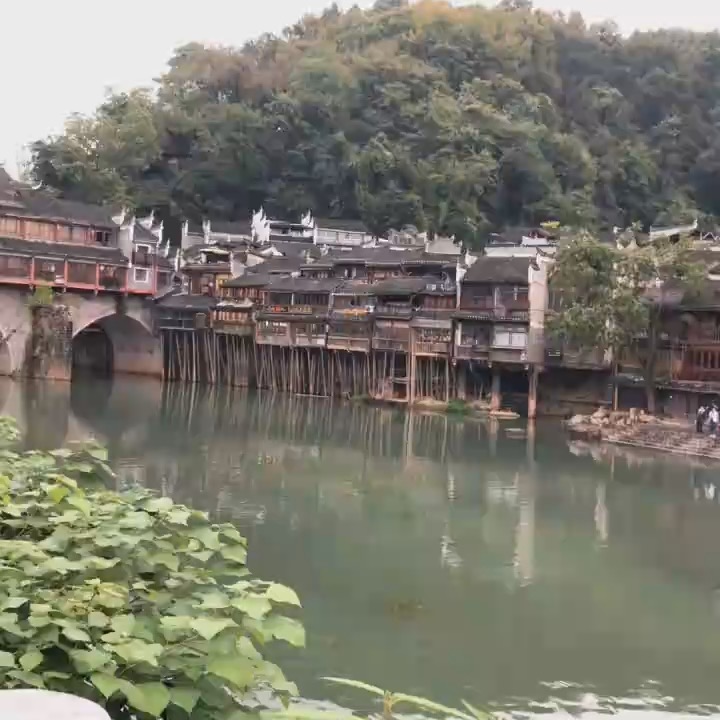}%
\label{fig:fr_map_dist_3}}
\hfil
\subfloat[]{\includegraphics[width=0.12\textwidth]{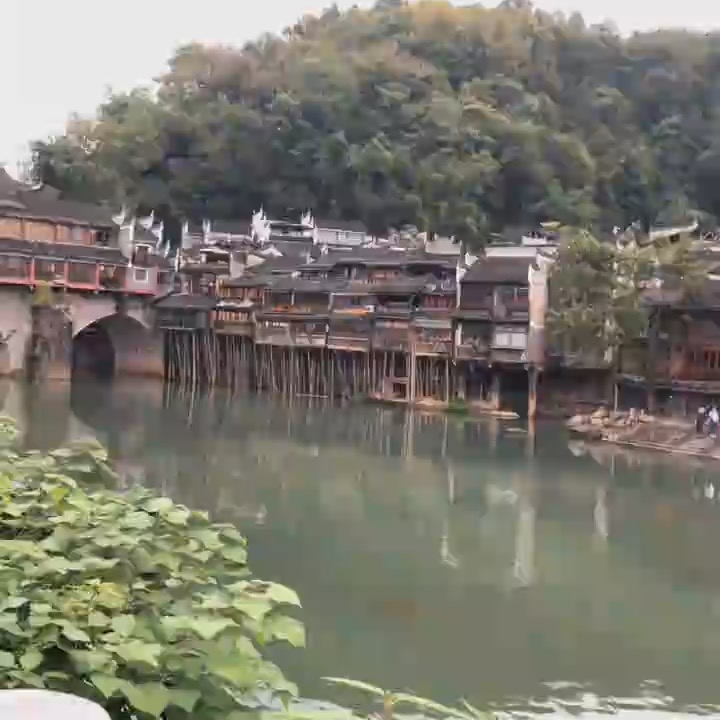}%
\label{fig:fr_map_dist_4}}
\hfil
\subfloat[]{\includegraphics[width=0.12\textwidth]{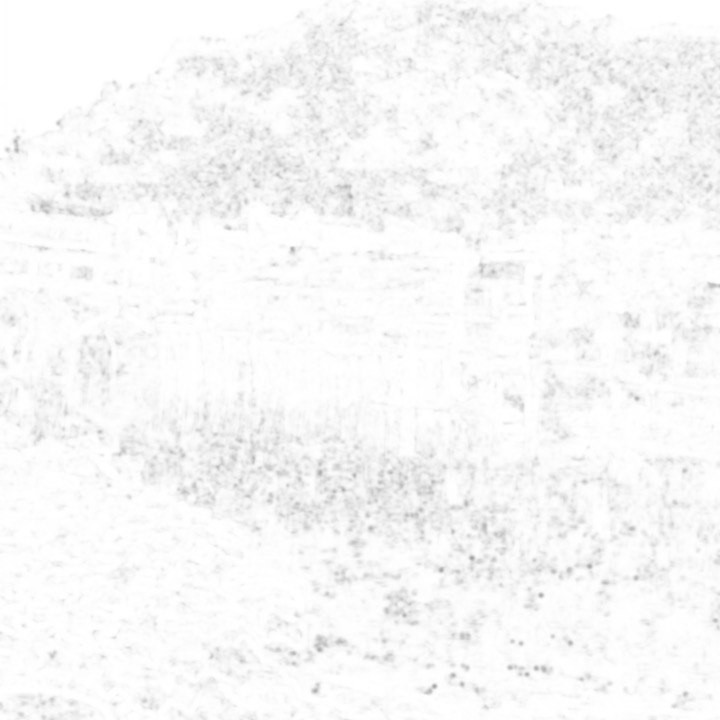}%
\label{fig:fr_map_ssim_1}}
\hfil
\subfloat[]{\includegraphics[width=0.12\textwidth]{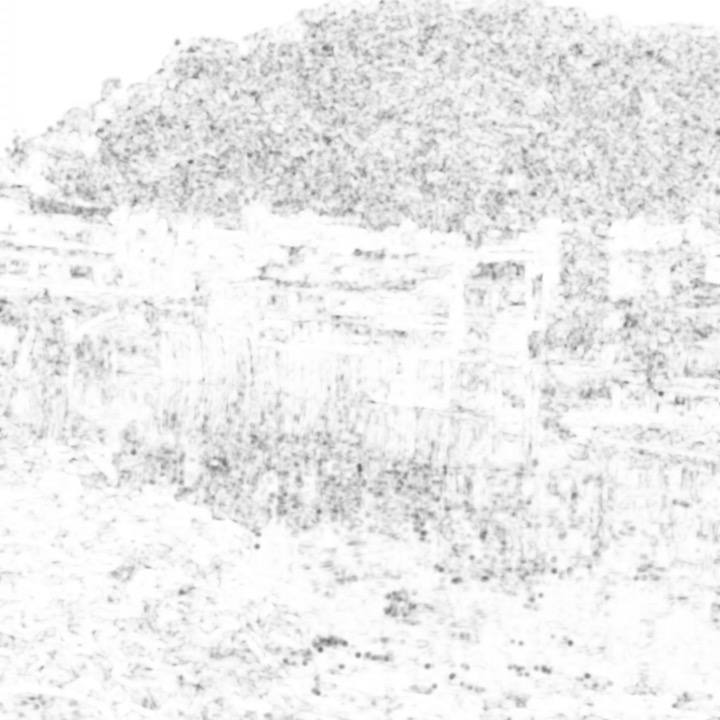}%
\label{fig:fr_map_ssim_2}}
\hfil
\subfloat[]{\includegraphics[width=0.12\textwidth]{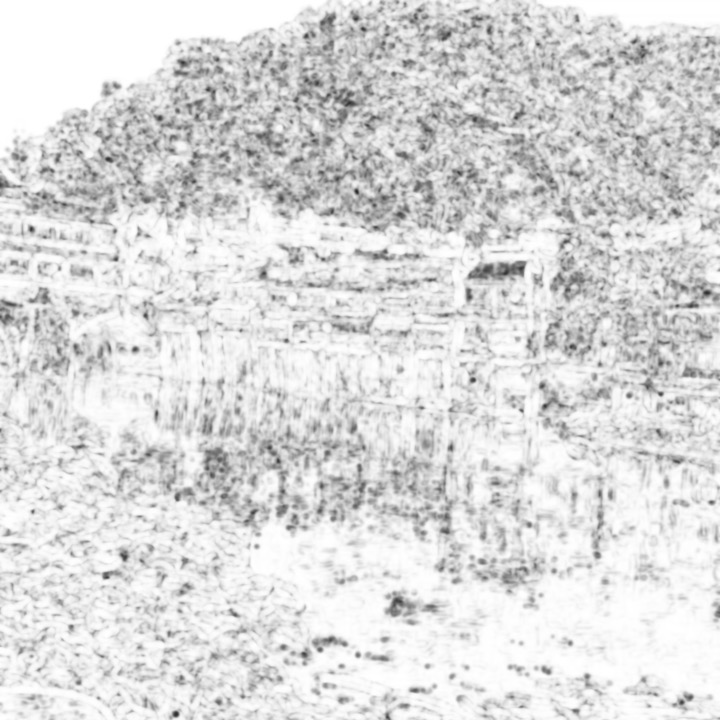}%
\label{fig:fr_map_ssim_3}}
\hfil
\subfloat[]{\includegraphics[width=0.12\textwidth]{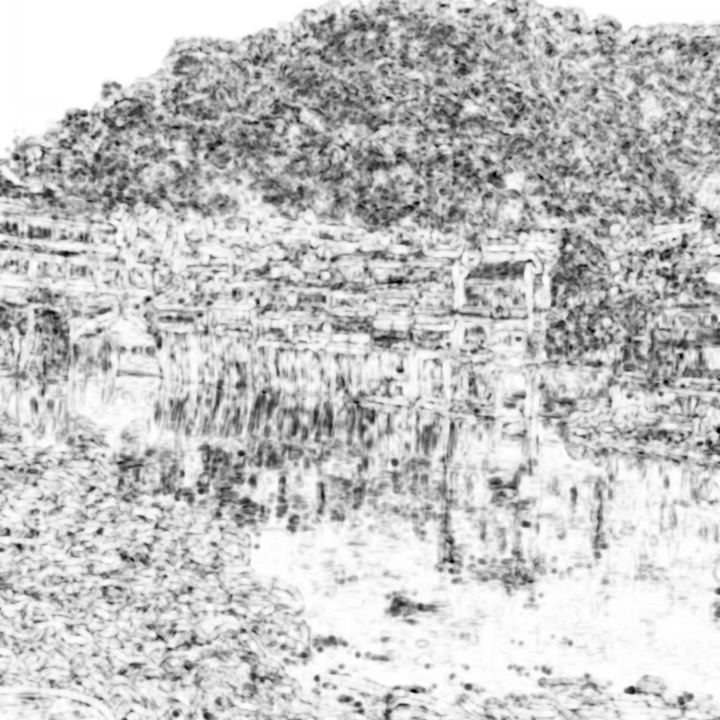}%
\label{fig:fr_map_ssim_4}}
\hfil
\subfloat[]{\includegraphics[width=0.12\textwidth]{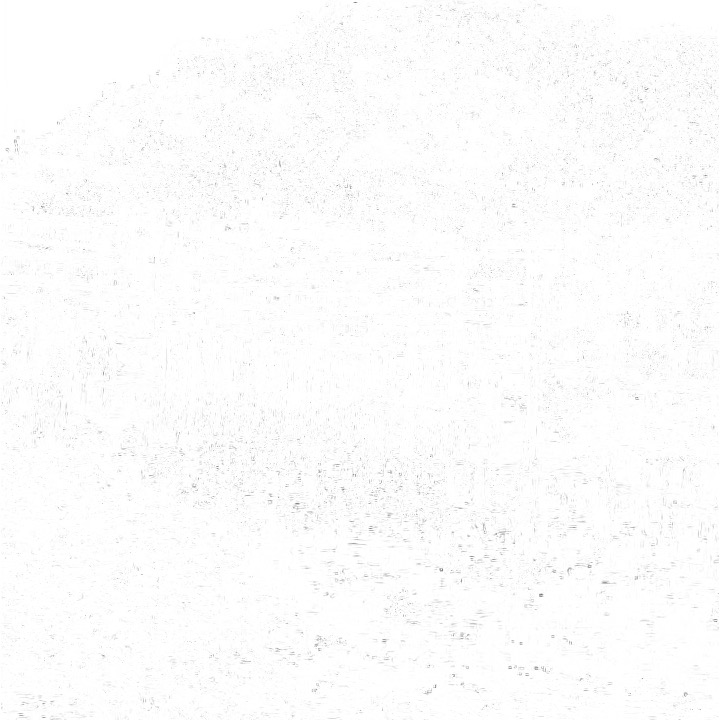}%
\label{fig:fr_map_mdsi_1}}
\hfil
\subfloat[]{\includegraphics[width=0.12\textwidth]{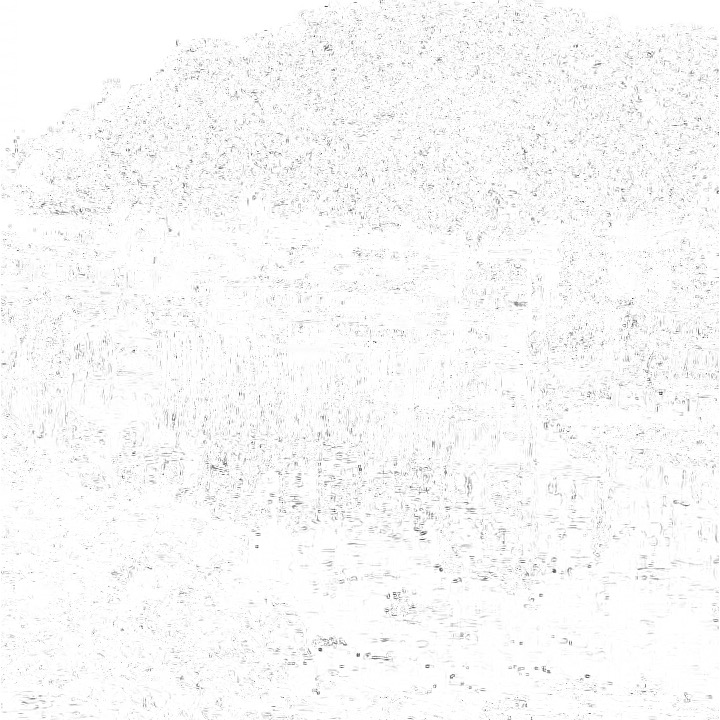}%
\label{fig:fr_map_mdsi_2}}
\hfil
\subfloat[]{\includegraphics[width=0.12\textwidth]{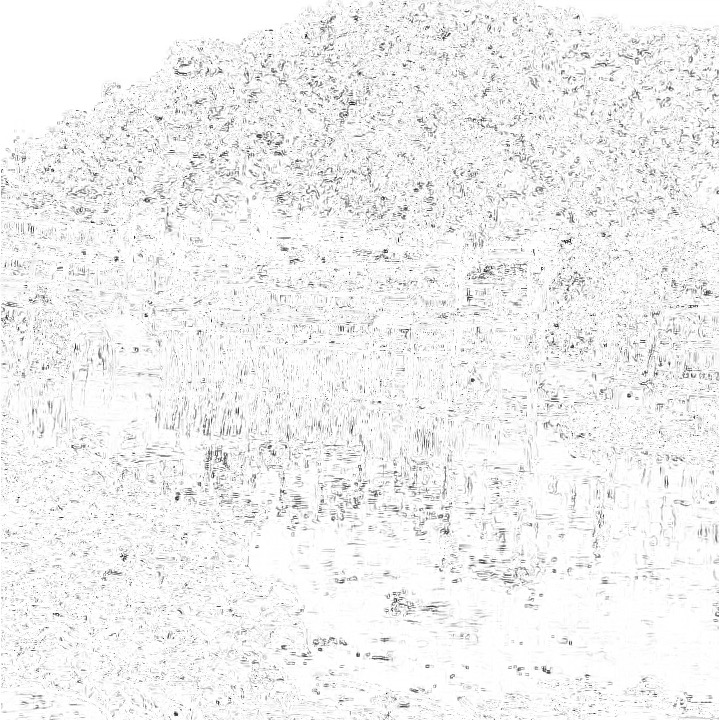}%
\label{fig:fr_map_mdsi_3}}
\hfil
\subfloat[]{\includegraphics[width=0.12\textwidth]{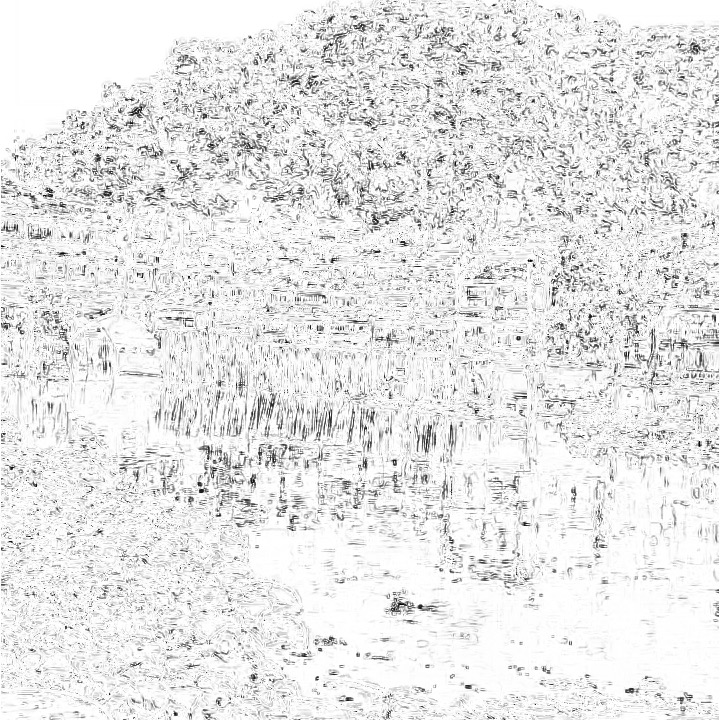}%
\label{fig:fr_map_mdsi_4}}
\hfil
\subfloat[]{\includegraphics[width=0.12\textwidth]{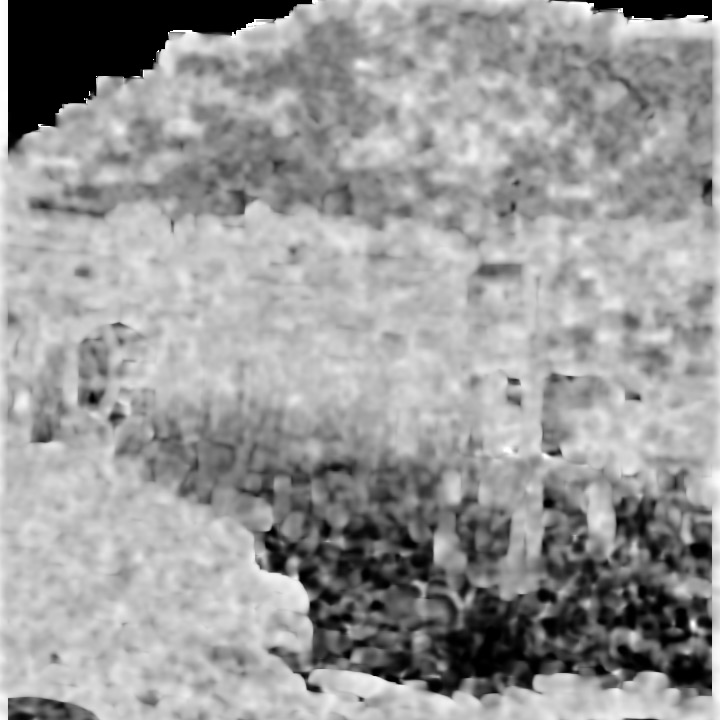}%
\label{fig:fr_map_vif_1}}
\hfil
\subfloat[]{\includegraphics[width=0.12\textwidth]{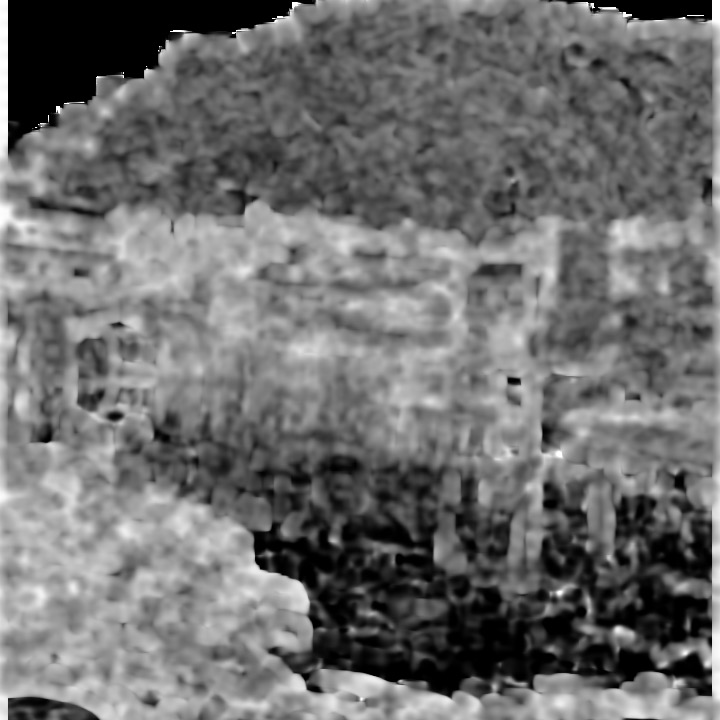}%
\label{fig:fr_map_vif_2}}
\hfil
\subfloat[]{\includegraphics[width=0.12\textwidth]{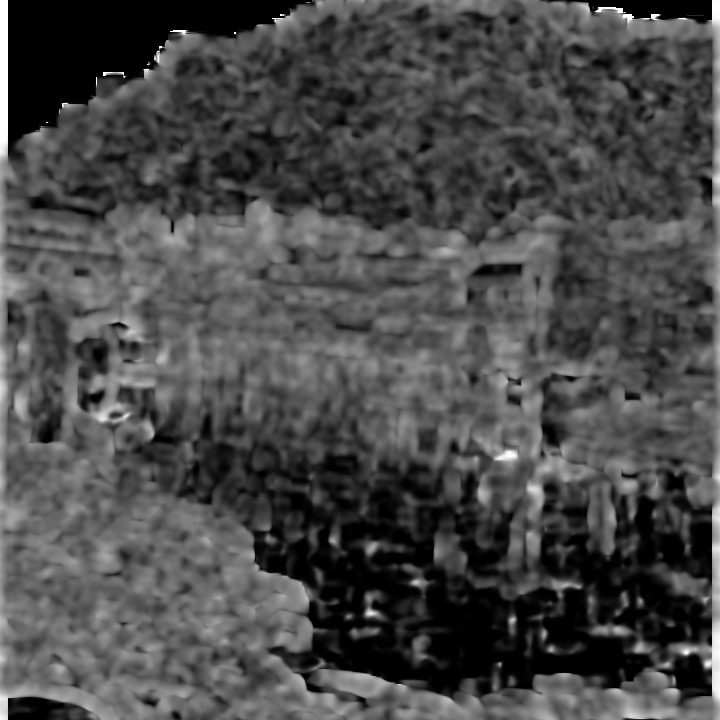}%
\label{fig:fr_map_vif_3}}
\hfil
\subfloat[]{\includegraphics[width=0.12\textwidth]{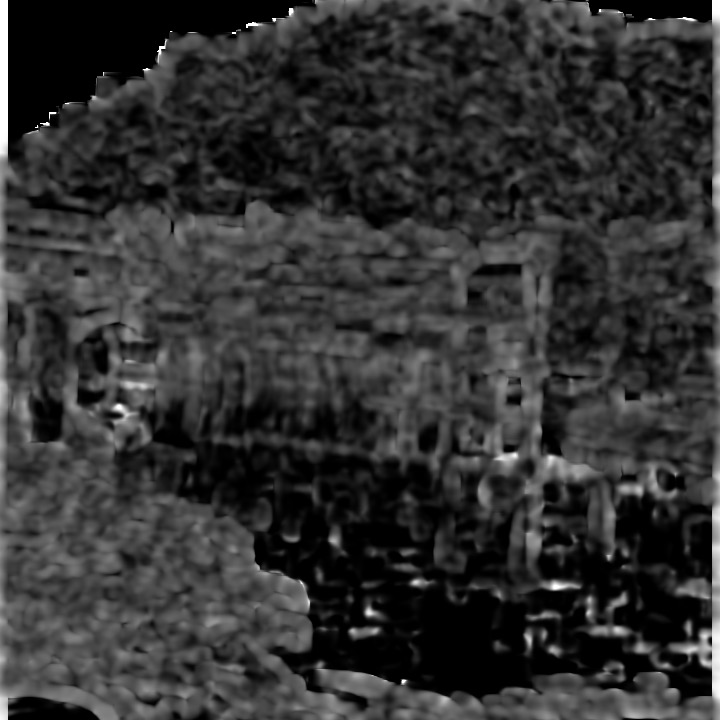}%
\label{fig:fr_map_vif_4}}
\caption{Quality maps generated based on $\bm{V}_s$ and $\bm{V}_c$. \protect\subref{fig:fr_map_ref} One frame from $\bm{V}_s$; (b)-(e) HEVC transcoded versions from $\bm{V}_s$ with QP 27, 32, 37 and 42; (f)-(i) corresponding SSIM maps; (j)-(m) corresponding MSDI maps; (n)-(q) corresponding VIF maps.}
\label{Fig:qualitymap}
\end{figure*}

\section{Objective Quality Assessment}
\label{sec:oqa}

\subsection{Framework}
\label{ssec:framework}

As illustrated in Fig.~\ref{Fig:framework}, the proposed UGC video quality assessment framework leverages the intrinsic quality of the source videos as well as the comparisons between the source and transcoded videos. Instead of straightforwardly obtaining a quality score of the source videos using NR algorithms, we propose to learn and fuse the intermediate quality maps, which meaningfully indicate the spatially variant quality of different regions. 
The inferred quality maps are fed to a pooling network, such that the local distortions are aggregated in a data-driven manner for final quality prediction. 
As such, the proposed quality evaluation framework consists of three main modules, including a generative network $G_{\phi}$ that generates the quality maps of the source videos, an evaluator $E_{\omega}$ which produces the relative quality maps between the source and transcoded videos, and a regression network $f_{\theta}$ that fuses the quality maps to obtain the final quality score. These modules are parameterized by $\phi$, $\omega$ and $\theta$, respectively.


Given a source video $\bm{V}_s$ and its transcoded version $\bm{V}_t$, we first predict the quality maps $\bm{M}_s$ of $\bm{V}_s$ using $G_{\phi}$: 
\begin{equation}
\label{eq:gennet}
M_s^i=G_{\phi}(I_s^i)
\end{equation}
where $I_s^i$ is the $i$-th frame of $\bm{V}_s$ and $M_s^i$ represents its corresponding quality map. 
Meanwhile the relative perceptual degradation between $\bm{V}_s$ and $\bm{V}_t$ can also be measured,
\begin{equation}
\label{eq:evaluator}
M_t^i=E_{\omega}(I_s^i, I_t^i)
\end{equation}
where $I_t^i$ and $M_t^i$ are the $i$-th frame of $\bm{V}_t$ and its corresponding quality map, respectively.
Finally, a quality pooling network $f_{\theta}$ concats the quality maps, which deliver the intrinsic quality of source videos as well as the relative quality between source and transcoded videos:
\begin{equation}
\label{eq:frmawork}
\hat{S}=f_{\theta}(G_{\phi}(\bm{V}_s), E_{\omega}(\bm{V}_s, \bm{V}_t))
\end{equation}
where $\hat{S}$ is the predicted score of transcoded video.

\subsection{Quality Maps Generation Based on  $\bm{V}_s$ and $\bm{V}_t$ }
\label{ssec:frqualitymap}

\begin{figure*}[ht]
\centering
\includegraphics[width=1.0\textwidth]{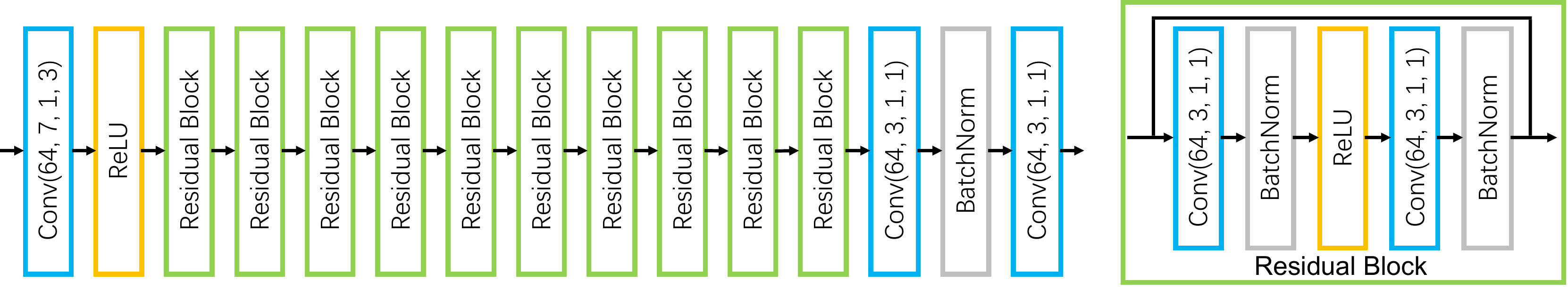}
\caption{The architecture of the generative network that produces the quality maps from  $\bm{V}_s$. Blue box: a convolutional layer $Conv(d,f,s,p)$ with $d$ filters of size $f\times f$, a stride of $s$ and a padding of $p$; yellow box: ReLU layer; gray box: batch normalization layer.}
\label{Fig:gennet}
\end{figure*}

\begin{figure*}[ht]
\centering
\subfloat[]{\includegraphics[width=0.14\textwidth]{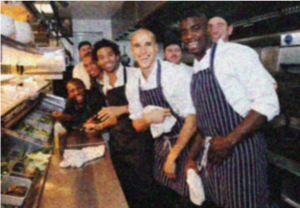}%
\label{fig:nr_1_dist}}
\hfil
\subfloat[]{\includegraphics[width=0.14\textwidth]{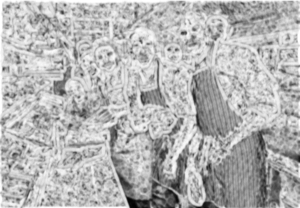}%
\label{fig:nr_1_ssim}}
\hfil
\subfloat[]{\includegraphics[width=0.14\textwidth]{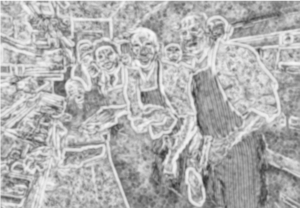}%
\label{fig:nr_1_ssim_p}}
\hfil
\subfloat[]{\includegraphics[width=0.14\textwidth]{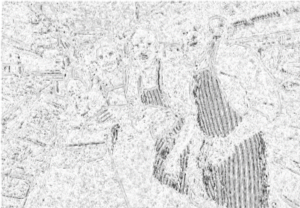}%
\label{fig:nr_1_mdsi}}
\hfil
\subfloat[]{\includegraphics[width=0.14\textwidth]{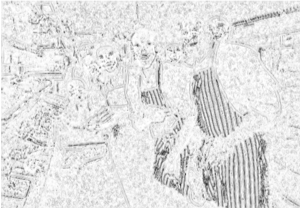}%
\label{fig:nr_1_mdsi_p}}
\hfil
\subfloat[]{\includegraphics[width=0.14\textwidth]{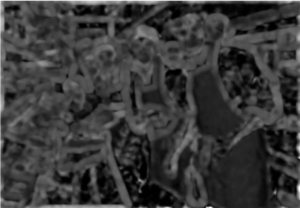}%
\label{fig:nr_1_vif}}
\hfil
\subfloat[]{\includegraphics[width=0.14\textwidth]{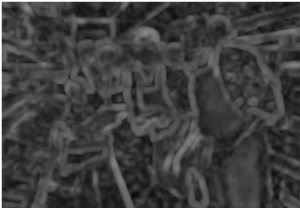}%
\label{fig:nr_1_vif_p}}
\hfil
\subfloat[]{\includegraphics[width=0.14\textwidth]{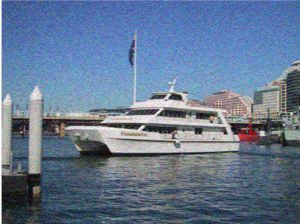}%
\label{fig:nr_2_dist}}
\hfil
\subfloat[]{\includegraphics[width=0.14\textwidth]{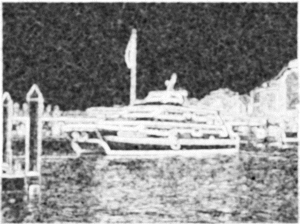}%
\label{fig:nr_2_ssim}}
\hfil
\subfloat[]{\includegraphics[width=0.14\textwidth]{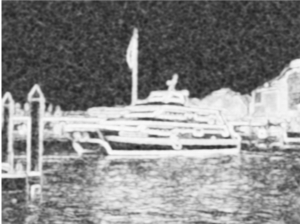}%
\label{fig:nr_2_ssim_p}}
\hfil
\subfloat[]{\includegraphics[width=0.14\textwidth]{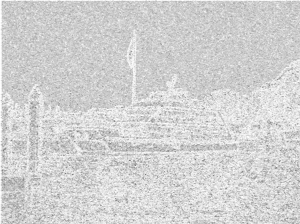}%
\label{fig:nr_2_mdsi}}
\hfil
\subfloat[]{\includegraphics[width=0.14\textwidth]{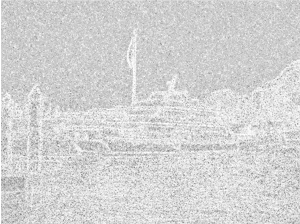}%
\label{fig:nr_2_mdsi_p}}
\hfil
\subfloat[]{\includegraphics[width=0.14\textwidth]{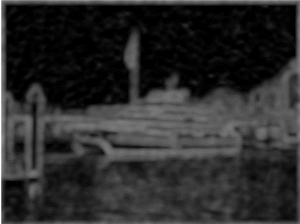}%
\label{fig:nr_2_vif}}
\hfil
\subfloat[]{\includegraphics[width=0.14\textwidth]{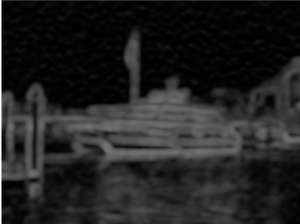}%
\label{fig:nr_2_vif_p}}
\caption{Illustration of the predicted quality maps and the corresponding ground-truth maps. \protect\subref{fig:nr_1_dist}\protect\subref{fig:nr_2_dist} distorted images with multiple distortions; \protect\subref{fig:nr_1_ssim}\protect\subref{fig:nr_2_ssim} ground-truth SSIM maps; \protect\subref{fig:nr_1_ssim_p}\protect\subref{fig:nr_2_ssim_p} predicted SSIM maps; \protect\subref{fig:nr_1_mdsi}\protect\subref{fig:nr_2_mdsi} ground-truth MDSI maps; \protect\subref{fig:nr_1_mdsi_p}\protect\subref{fig:nr_2_mdsi_p} predicted MDSI maps; \protect\subref{fig:nr_1_vif}\protect\subref{fig:nr_2_vif} ground-truth VIF maps; \protect\subref{fig:nr_1_vif_p}\protect\subref{fig:nr_2_vif_p} predicted VIF maps.}
\label{Fig:predictedmap}
\end{figure*}

\begin{figure*}[ht]
\centering
\subfloat[]{\includegraphics[width=0.12\textwidth]{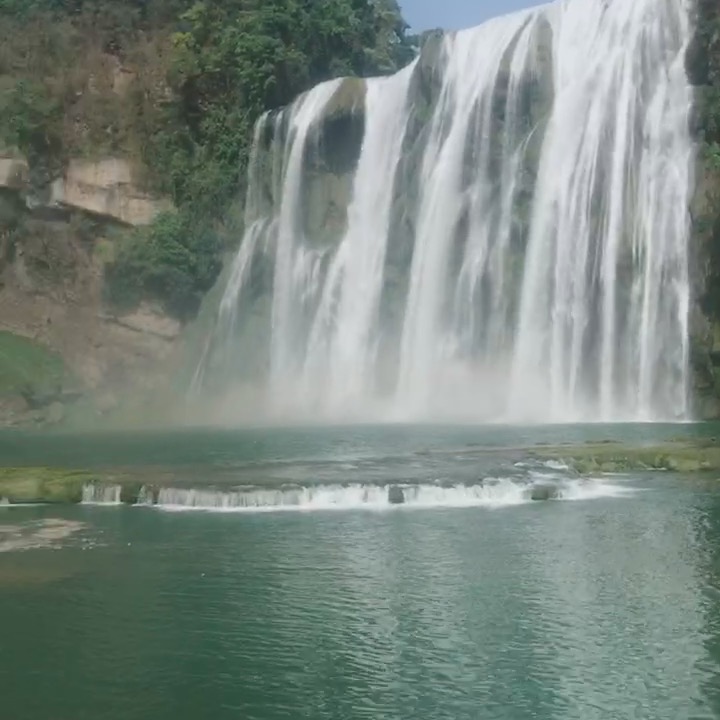}%
\label{fig:p_nr_1_dist}}
\hfil
\subfloat[]{\includegraphics[width=0.12\textwidth]{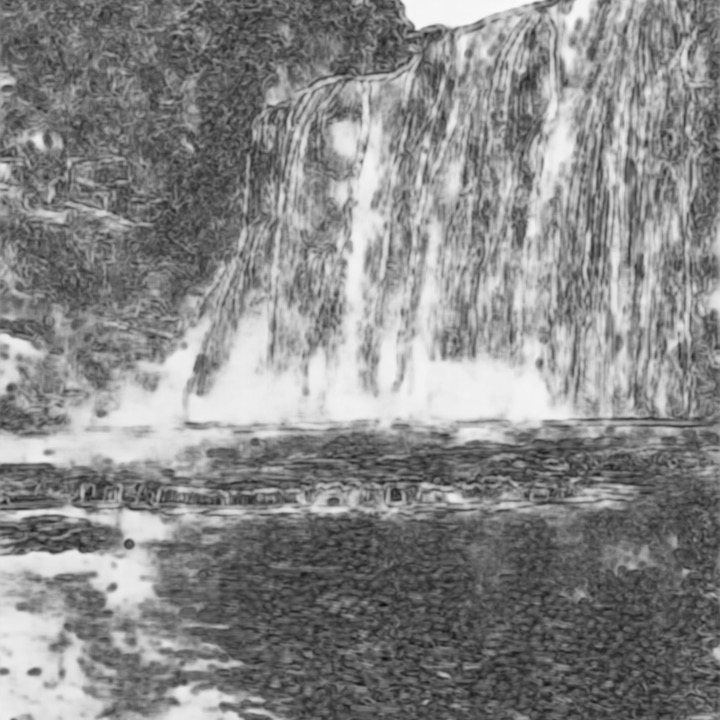}%
\label{fig:p_nr_1_ssim}}
\hfil
\subfloat[]{\includegraphics[width=0.12\textwidth]{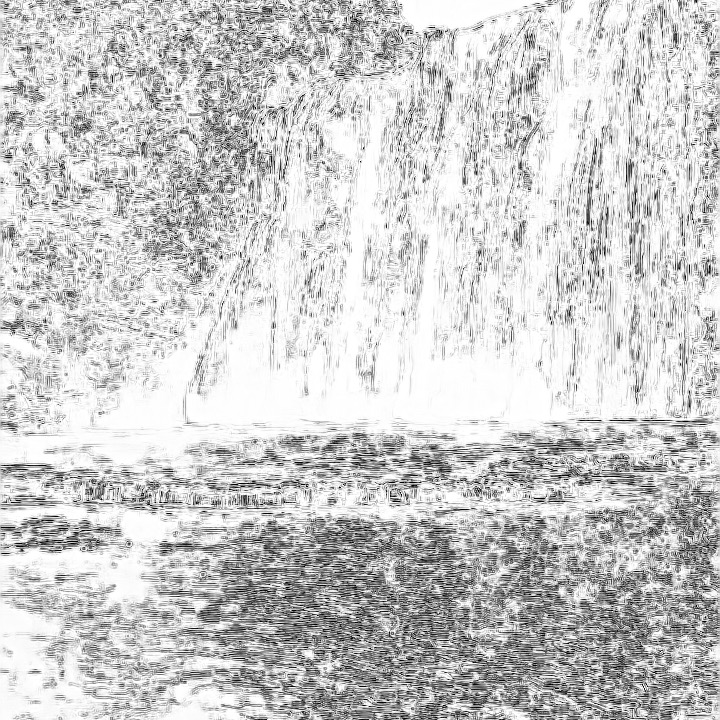}%
\label{fig:p_nr_1_mdsi}}
\hfil
\subfloat[]{\includegraphics[width=0.12\textwidth]{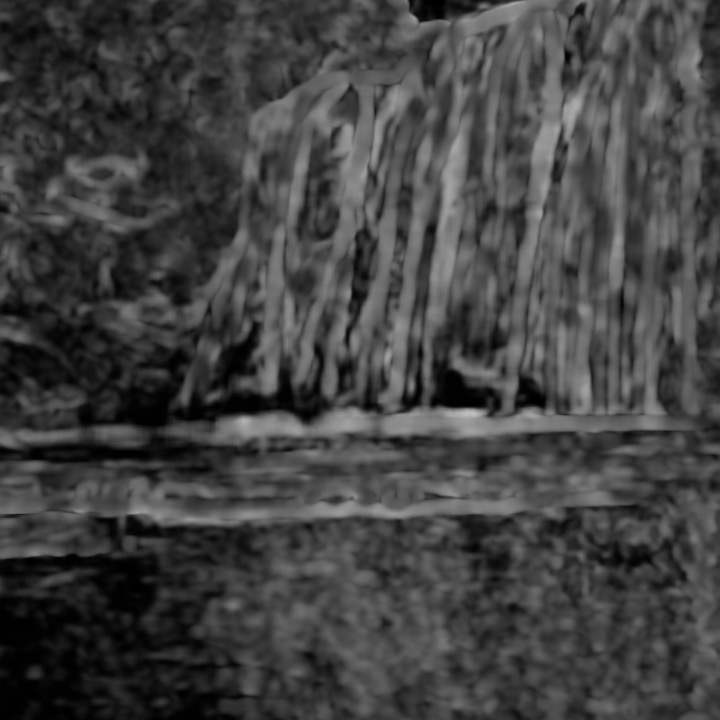}%
\label{fig:p_nr_1_vif}}
\hfil
\subfloat[]{\includegraphics[width=0.12\textwidth]{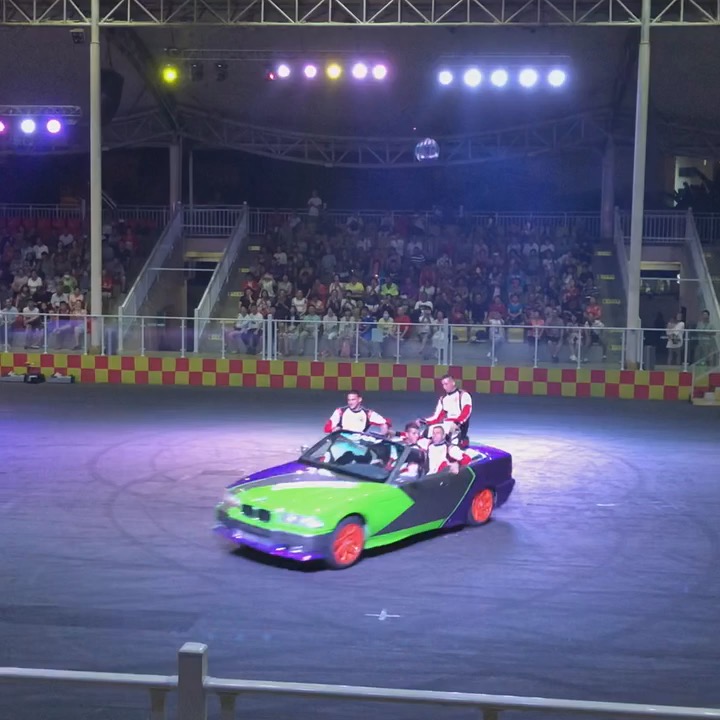}%
\label{fig:p_nr_2_dist}}
\hfil
\subfloat[]{\includegraphics[width=0.12\textwidth]{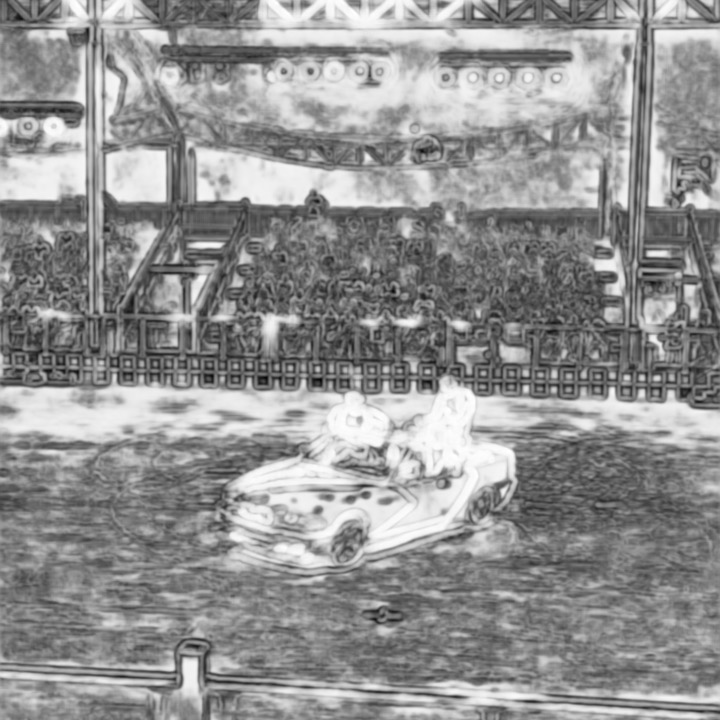}%
\label{fig:p_nr_2_ssim}}
\hfil
\subfloat[]{\includegraphics[width=0.12\textwidth]{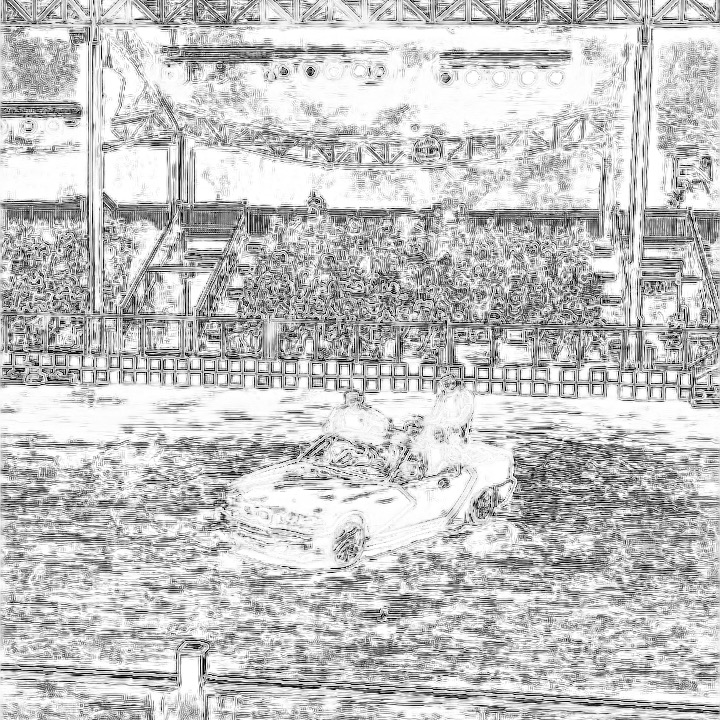}%
\label{fig:p_nr_2_mdsi}}
\hfil
\subfloat[]{\includegraphics[width=0.12\textwidth]{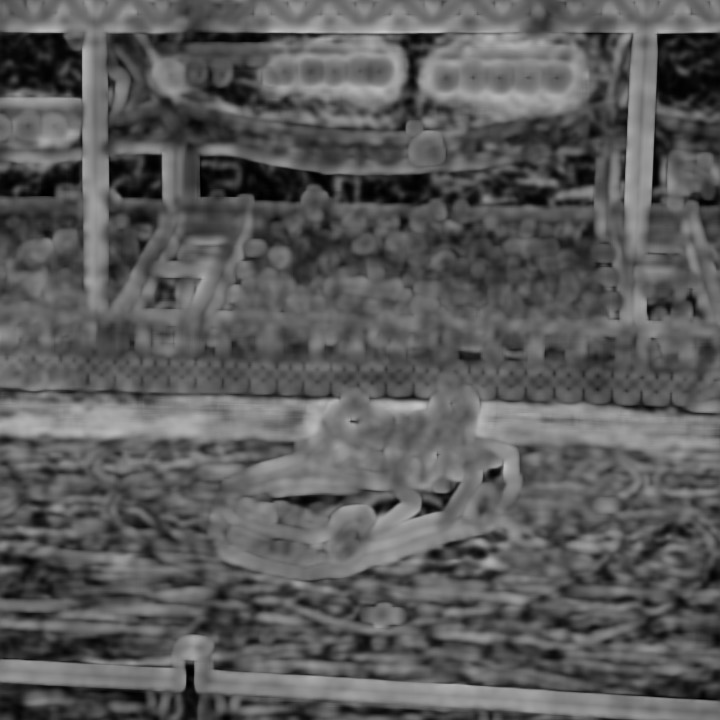}%
\label{fig:p_nr_2_vif}}
\caption{Illustration of the predicted quality maps in our database. \protect\subref{fig:p_nr_1_dist}\protect\subref{fig:p_nr_2_dist} frames of $\bm{V}_s$; \protect\subref{fig:p_nr_1_ssim}\protect\subref{fig:p_nr_2_ssim} predicted SSIM maps of $\bm{V}_s$; \protect\subref{fig:p_nr_1_mdsi}\protect\subref{fig:p_nr_2_mdsi} predicted MDSI maps of $\bm{V}_s$; \protect\subref{fig:p_nr_1_vif}\protect\subref{fig:p_nr_2_vif} predicted VIF maps of $\bm{V}_s$.}
\label{Fig:predictedmaponugc}
\end{figure*}

In the hosting platform, $\bm{V}_s$ is further transcoded into $\bm{V}_t$, such that the difference lying between them originates from compression artifacts. As such, given  $\bm{V}_s$ and $\bm{V}_t$, to evaluate the relative distortion between them, we leverage existing quality metrics including SSIM~\cite{SSIM}, MDSI~\cite{MDSI} and VIF~\cite{VIF}, which well reflect the local distortion from $I_s^i$ to $I_t^i$ from the perspectives of structure, gradient and visual information, respectively. Regarding SSIM, only luminance component is considered and the derived single channel luminance similarity map of each frame pair is used as SSIM quality map. 
With respect to MDSI, the combination of gradient similarity map and chromaticity similarity map is used as MDSI map.
Since VIF is a multi-scale method, only VIF map derived from the frames of the original size is adopted.
These quality maps are shown in Fig.~\ref{Fig:qualitymap}, which imply that the adopted quality maps well predict the visual quality. 
The values in the quality maps are normalized to the range of $\left[0, 1 \right]$ to facilitate subsequent training in DNN.


\subsection{Quality Maps Generation from Source Video $\bm{V}_s$}
\label{ssec:nrqualitymap}

Given the source video  $\bm{V}_s$, we aim at blindly estimating the quality map of each frame since the pristine reference is not available. 
We adopt the deep neural networks ensuring the robust and accurate quality map prediction. In particular, ResNet~\cite{he2016deep} is employed with the consideration that residual connections make the training of identical function easier, which gradually facilitate the adding of distortions from low level to high level.
The detailed architecture of the generative network is shown in Fig.~\ref{Fig:gennet}. More specifically, quality maps of the input frame are predicted after 10 identical residual blocks, each of which contains two 3$\times$3 conventional layers with 64 feature maps, and all convolution layers are with stride 1$\times$1 and zero-padding. As such, the size of the final output feature map is consistent with the original input frame. 
Besides, batch normalization~\cite{ioffe2015batch} and rectified linear unit (ReLU) are used after convolution.

\begin{figure*}[ht]
\centering
\includegraphics[width=0.9\textwidth]{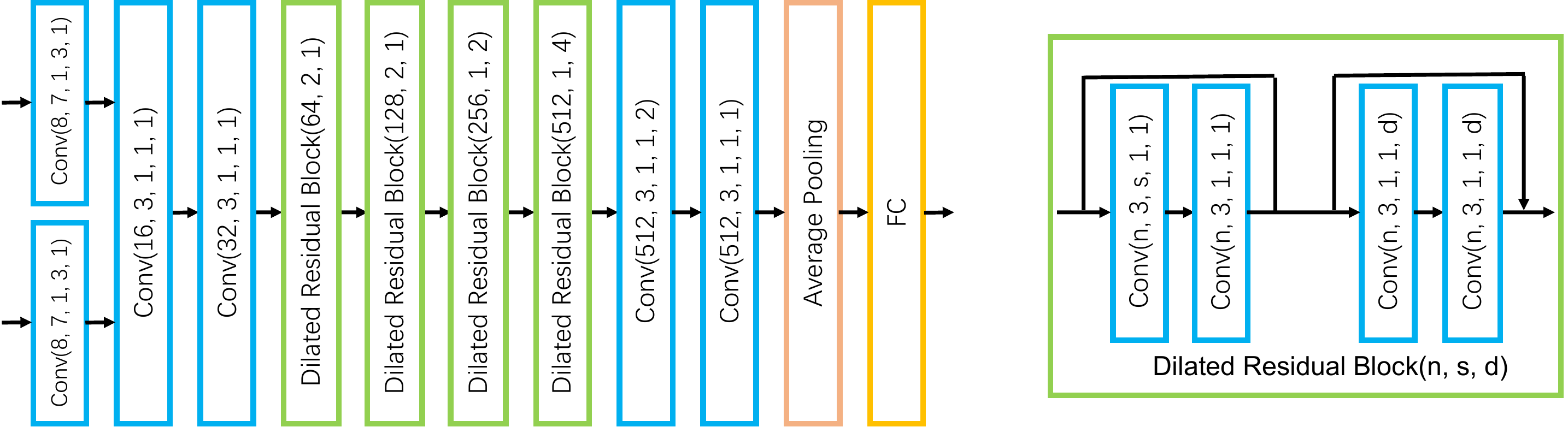}
\caption{Detailed architecture of the pooling network. Blue box: a convolutional layer $Conv(n,f,s,p,d)$ with $n$ filters of size $f\times f$, a stride of $s$, a padding of $p$ and a dilation of $d$; red box: average pooling layer; yellow box: full connection layer. It is worth mentioning that there are batch normalization and ReLU layers after each convolutional layer, which are omitted here for simplification.}
\label{Fig:poolnet}
\end{figure*}

Regarding the training of the quality map generative network, Waterloo database~\cite{ma2016waterloo} is adopted. In particular, it includes 4744 pristine images, as well as distorted versions with Gaussian noise, Gaussian blur, JPEG compression and JPEG2000 compression.
It is worth mentioning that we use the pristine images in the database to regenerate the distorted images. To model the distortions contained in $\bm{V}_s$, multiple distortion stages are applied on these pristine images.
Gaussian blur or Gaussian noise of different levels is injected to the pristine image, and subsequently these distorted images are compressed with certain compression levels by JPEG or JPEG2000 compression. The images after compression are used as training inputs and their quality maps are used as groundtruth labels. As described in Section~\ref{ssec:frqualitymap}, different quality maps derived from existing FR methods can be adopted as training labels. Different quality maps predicted by the generative network and their corresponding ground-truth labels are shown in Fig.~\ref{Fig:predictedmap}. 
The generative networks trained on Waterloo database are then applied on our database to generate quality maps of $\bm{V}_s$, as shown in Fig.~\ref{Fig:predictedmaponugc}.

The loss function for the generative network consists of a structural loss characterized by SSIM  and pixel-wise loss, as introduced in~\cite{zhao2016loss}, which is given by,
\begin{equation}
\label{eq:genloss}
L_{G}(P_k^0, P_k)=\alpha \cdot L^{SSIM}(P_k^0, P_k) + (1-\alpha) \cdot L^{L1}(P_k^0, P_k)
\end{equation}
where $P_k^0$ is the ground-truth quality map patch, $P_k$ is the corresponding generated patch, and $\alpha$ here is an empirically set weighting factor.
The structural loss based on SSIM is formulated as, 
\begin{equation}
\label{eq:ssimloss}
L^{SSIM}(P_k^0, P_k) = 1 - SSIM(P_k^0, P_k).
\end{equation}

\subsection{Quality Map Pooling}
\label{ssec:poolingnetwork}

After the generation of quality maps from the source video and transcoded video, a pooling network is trained to fuse these quality maps and generate a final quality score. 
In general, convolutional networks have been widely used to progressively reduce the resolution of feature maps, while such loss of spatial acuity may limit the performance.
In our framework, a dilated residual network (DRN)~\cite{yu2017dilated} is employed, in which dilated convolutions are used to increase the resolution of output feature maps without reducing the receptive field of individual neurons.
As shown in Fig.~\ref{Fig:poolnet}, each set of the input maps flows through independent convolutional layers, and feature maps are concatenated after the first convolutional layer.
The four dilated residual structures with 3$\times$3 convolution kernel are used to extract feature representations, and average pooling as well as fully connected layers are used to regress to the final score.
The loss function of pooling network is mean square error (MSE) loss, 
\begin{equation}
\label{eq:regloss}
L_{REG}=\left\lVert f_{\theta}(G_{\phi}(I_s^i), E_{\omega}(I_s^i, I_t^i))-S  \right\lVert^2
\end{equation}
where $S$ denote human evaluation for $\bm{V}_t$, and the video score is set as training label for each quality map pairs.
Scores of all sampled frames are pooled at the sequence level by average pooling.
In this step, 30 frames are uniformly sampled from each video to train the model.

\section{Experimental Results}
\label{sec:experiment}


\subsection{Experimental Settings}
\label{ssec:trainingsetting}

\subsubsection{Database}
\label{sssec:experimentdatabase}

Due to the lack of databases that align with the UGC application scenario, in particular from acquisition to processing on the hosting platform, the newly introduced UGC-Video database in Section~\ref{sec:database} is used in evaluating our proposed method.

\subsubsection{Compared methods}
\label{sssec:experimentmethods}

Both FR and NR quality assessment algorithms are applicable for quality assessment of UGC videos. In particular, in FR methods, source videos with various quality levels are used as reference. The NR methods can be directly applied on the transcoded videos for evaluating the quality of compressed videos. 
Effective FR and NR methods with high generalization capaiblity are used for comparison, including PSNR, SSIM, MS-SSIM, VIF, NIQE, BRISQUE, VBLIINDS, VIIDEO, VMAF. 
In addition, a 2stepQA~\cite{yu2019predicting} method combining FR and NR models are also considered which serves as a flexible framework based on different combinations of FR and NR methods.

\subsection{Training Details}
\label{ssec:rainingdetail}
The training process consists of two steps: (1) training generative network on the modified Waterloo Exploration Database; (2) training the pooling network on UGC-Videos. 

In the original Waterloo Exploration Database~\cite{ma2016waterloo}, 94880 distorted images are created from 4744 pristine natural images by introducing four types of distortion (blur, noise, JPEG and JPEG2K), each with five levels.
To enable the generation network to capture mixture distortions similar to that in the source UGC videos, we develop a new way to generate the distorted images.
More specifically, noise or blur distortions of random level are first induced to these pristine images, and subsequently compression distortion is also injected by JPEG or JPEG2000 with random compression level. As such, 4744 distorted images with multiple distortions are created.
VIF quality maps are calculated according the distorted images and the corresponding pristine images.
Both distorted images and their quality maps are cropped into 64$\times$64 non-overlapping patches. Generative network is trained based on the inputs (patches from distorted image) and labels (corresponding patches from VIF quality map) using Adam optimizer~\cite{kingma2014adam} at the initial learning rate of $10^{-3}$ for 100 epochs.


Subsequently, the pooling network is trained using the pre-trained generative network model and  the score is regressed using quality maps.
Once each quality map is derived from the previous generative network, we freeze the weights of generative networks, and train the pooling network using MSE loss and Adam optimizer with the learning rate of $10^{-4}$. By the dilated residual blocks and average pooling, quality map pairs with the fully connected layers yield the final score.

\subsection{Performance Comparisons}
\label{ssec:performancecomparison}

\begin{table}[t]
\caption{Mean and standard deviation of performance values of various FR and NR methods in 20 runs on UGC-Video database, i.e., mean ($\pm$std)}
\label{tab:exp_r_nr}
\centering
\begin{tabular}{|c|c|c|c|}
\hline
Method                  & SROCC              & PLCC               & RMSE              \\ \hline
PSNR                    & 0.647 ($\pm$0.098) & 0.659 ($\pm$0.083) & 0.649 ($\pm$0.046) \\ \hline
SSIM                    & 0.729 ($\pm$0.102) & 0.778 ($\pm$0.088) & 0.536 ($\pm$0.092) \\ \hline
MS-SSIM                 & 0.735 ($\pm$0.095) & 0.782 ($\pm$0.086) & 0.782 ($\pm$0.086) \\ \hline
VIF                     & 0.770 ($\pm$0.067) & 0.737 ($\pm$0.113) & 0.602 ($\pm$0.137) \\ \hline
VideoSpEED              & 0.790 ($\pm$0.074) & 0.810 ($\pm$0.079) & 0.499 ($\pm$0.103) \\ \hline
ViS3                    & 0.764 ($\pm$0.082) & 0.788 ($\pm$0.085) & 0.527 ($\pm$0.091) \\ \hline
VMAF                    & 0.831 ($\pm$0.044) & 0.875 ($\pm$0.043) & 0.416 ($\pm$0.074) \\ \hline
NIQE                    & 0.338 ($\pm$0.102) & 0.275 ($\pm$0.103) & 0.813 ($\pm$0.045) \\ \hline
BRISQUE                 & 0.392 ($\pm$0.098) & 0.323 ($\pm$0.091) & 0.800 ($\pm$0.045) \\ \hline
VIIDEO                  & 0.110 ($\pm$0.080) & 0.133 ($\pm$0.064) & 0.835 ($\pm$0.040) \\ \hline
VBLINDS                 & 0.233 ($\pm$0.115) & 0.232 ($\pm$0.101) & 1.102 ($\pm$1.121) \\ \hline
\end{tabular}
\end{table}

\begin{table}[t]
\scriptsize
\caption{Mean and standard deviation of performance values of 2stepQA model using different combinations of FR and NR methods in 20 runs on UGC-Video database, i.e., mean ($\pm$std)}
\label{tab:exp_2stepqa}
\centering
\begin{tabular}{|c|c|c|c|}
\hline
Method        & SROCC              & PLCC               & RMSE               \\ \hline
PSNR+NIQE     & 0.687 ($\pm$0.102) & 0.699 ($\pm$0.108) & 0.613 ($\pm$0.079) \\ \hline
PSNR+BRISQUE  & 0.762 ($\pm$0.055) & 0.756 ($\pm$0.048) & 0.565 ($\pm$0.051) \\ \hline
PSNR+VIIDEO   & 0.636 ($\pm$0.096) & 0.654 ($\pm$0.087) & 0.653 ($\pm$0.051) \\ \hline
PSNR+VBLIINDS & 0.664 ($\pm$0.091) & 0.658 ($\pm$0.082) & 0.649 ($\pm$0.046) \\ \hline
SSIM+NIQE     & 0.734 ($\pm$0.091) & 0.781 ($\pm$0.088) & 0.529 ($\pm$0.096) \\ \hline
SSIM+BRISQUE  & 0.804 ($\pm$0.053) & 0.822 ($\pm$0.057) & 0.486 ($\pm$0.080) \\ \hline
SSIM+VIIDEO   & 0.727 ($\pm$0.103) & 0.777 ($\pm$0.088) & 0.538 ($\pm$0.093) \\ \hline
SSIM+VBLIINDS & 0.770 ($\pm$0.089) & 0.789 ($\pm$0.080) & 0.522 ($\pm$0.089) \\ \hline
VIF+NIQE      & 0.770 ($\pm$0.067) & 0.734 ($\pm$0.117) & 0.606 ($\pm$0.143) \\ \hline
VIF+BRISQUE   & 0.770 ($\pm$0.066) & 0.734 ($\pm$0.116) & 0.606 ($\pm$0.143) \\ \hline
VIF+VIIDEO    & 0.770 ($\pm$0.067) & 0.735 ($\pm$0.116) & 0.606 ($\pm$0.142) \\ \hline
VIF+VBLIINDS  & 0.764 ($\pm$0.067) & 0.727 ($\pm$0.124) & 0.622 ($\pm$0.178) \\ \hline
VMAF+NIQE     & 0.820 ($\pm$0.042) & 0.872 ($\pm$0.043) & 0.420 ($\pm$0.072) \\ \hline
VMAF+BRISQUE  & 0.846 ($\pm$0.031) & 0.886 ($\pm$0.033) & 0.398 ($\pm$0.065) \\ \hline
VMAF+VIIDEO   & 0.821 ($\pm$0.045) & 0.871 ($\pm$0.044) & 0.446 ($\pm$0.124) \\ \hline
VMAF+VBLIINDS & 0.829 ($\pm$0.035) & 0.874 ($\pm$0.038) & 0.417 ($\pm$0.067) \\ \hline
\end{tabular}
\end{table}

\begin{table}[t]
\caption{Mean and standard deviation of performance values of our proposed model using different combinations of quality maps, i.e., quality maps for transcoded video + quality maps for source video, in 20 runs on UGC-Video database. Mean and standard deviation (std) of performance values in 20 runs are reported, i.e., mean ($\pm$std)}
\label{tab:exp_ours}
\centering
\begin{tabular}{|c|c|c|c|}
\hline
Method   & SROCC              & PLCC               & RMSE               \\ \hline
SSIM map & 0.812 ($\pm$0.048) & 0.847 ($\pm$0.050) & 0.448 ($\pm$0.076) \\ \hline
VIF map  & 0.853 ($\pm$0.039) & 0.878 ($\pm$0.040) & 0.409 ($\pm$0.070) \\ \hline
VMAF map & 0.860 ($\pm$0.044) & 0.875 ($\pm$0.043) & 0.413 ($\pm$0.070) \\ \hline
\end{tabular}
\end{table}

\begin{figure}[t]
\centering
\includegraphics[width=1.0\columnwidth]{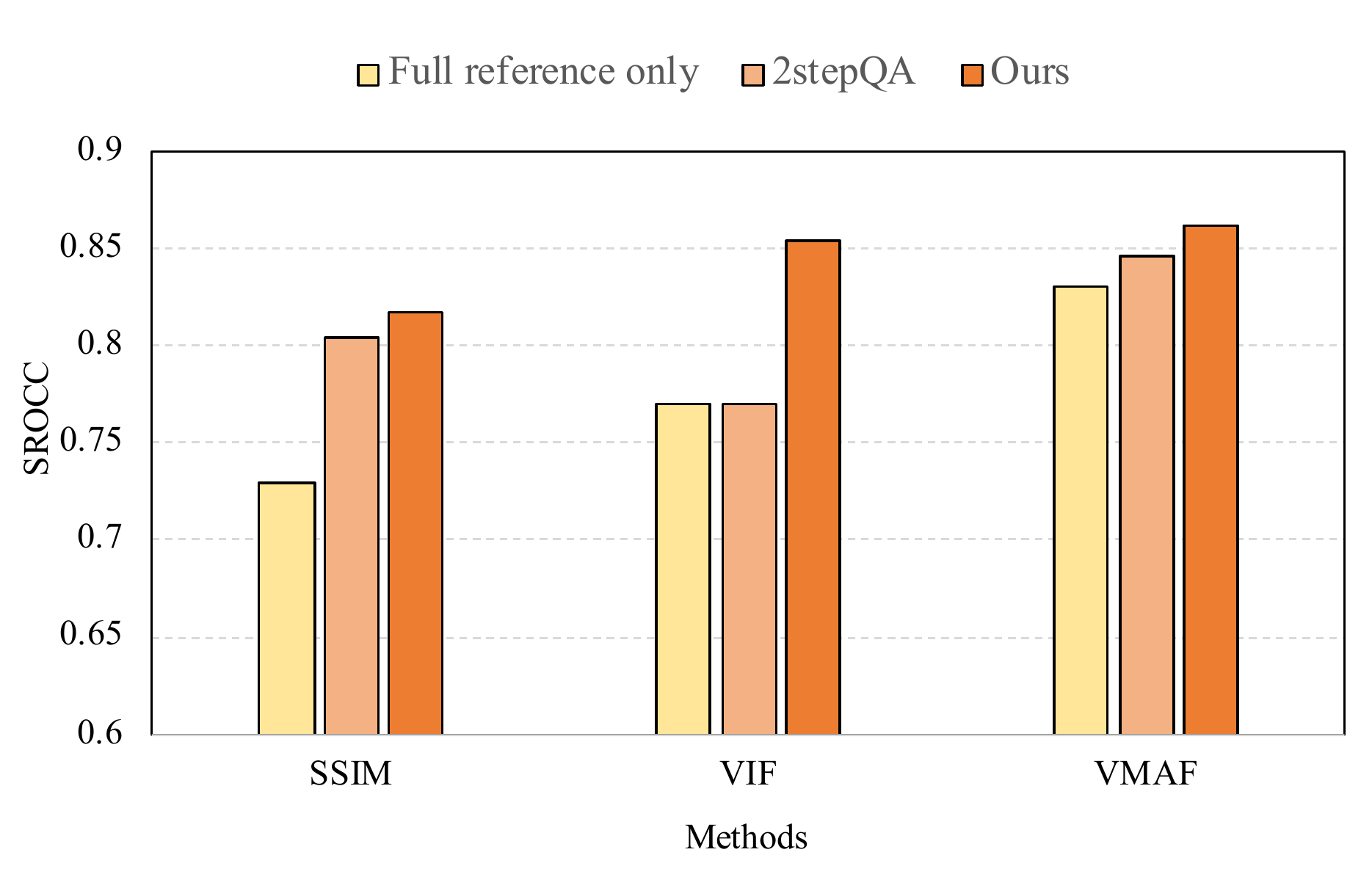}
\caption{SROCC performance of the compared algorithms over 20 trials on the UGC-Video database. }
\label{Fig:sroccbox}
\end{figure}

To ensure fair comparisons with existing conventional and learning-based methods, the full database is randomly divided into non-overlapping 60$\%$ training set, 20$\%$ validation set and 20$\%$ test set, according to the content of source videos.
Conventional quality measures which are not learning-based, i.e., PSNR, SSIM, VIF, NIQE, BRISQUE, VIIDEO and VBLIINDS are directly evaluated on the 20$\%$ testing data after the parameters in Eqn~\eqref{eqn:fitting} are optimized with the training and validation data.
For the 2stepQA method, the training and validation sets are merged together to train the relevant parameters.
For our method, the models with the highest SROCC value on the validation set during the training are chosen for testing.
This procedure has been repeated for 20 times and all above methods are tested on the same 20$\%$ test set. In particular, the mean and standard deviation of performance values are reported.

Table~\ref{tab:exp_r_nr} shows the performance of conventional methods, and it is apparent that FR algorithms perform better than NR algorithms, and the VMAF performs best by combining different metrics. 
Moreover, the 2stepQA performances with different combinations of FR and NR models are shown in Table~\ref{tab:exp_2stepqa}, 
we can see that the performances of reference algorithms have been improved in most cases. However, due to the simplicity of the 2stepQA model and the lack of efficient NR models with high generalization capability, 2stepQA method may degrade the performance of FR algorithms, such as VIF and VMAF.

The performances of the proposed framework are shown in Table~\ref{tab:exp_ours}, where predicted VIF quality maps by the generative network are obtained for the source video and different quality maps calculated using existing FR methods are used as the quality maps of transcoded videos. We can observe that our method is superior to the FR algorithms with more performance improvement compared with the 2stepQA method.
It is worth mentioning that VMAF quality map represents the concatenation of multiple types of quality maps, due to the fact that VMAF is a combination of multiple indicators.
More specifically, VIF quality map and motion map are contained in the VMAF quality map, where the motion map is luminance component difference calculated along the consecutive frames. 

To demonstrate the effectiveness of our framework, the average SROCC performance of FR methods, directly combining FR and NR scores (2stepQA) and our methods are compared in Fig.~\ref{Fig:sroccbox}. 
For SSIM method, both 2stepQA and our method greatly improve the accuracy of prediction, where 2stepQA increases SROCC from 0.729 to 0.804 by introducing the BRISQUE score of source video and our method increases the SROCC to 0.812 by combining the quality maps of source video and SSIM quality map of transcoded video. For VIF method, 2stepQA fails to improve the performance where our method brings great performance improvement.
As can be seen from Fig.~\ref{Fig:sroccbox}, our method significantly improves performance of existing reference models, and exhibit higher and more reliable correlations with subjective quality compared than the direct combination of the FR and NR algorithms.

\subsection{Ablation Studies}
\label{ssec:ablation}

\begin{table}[t]
\caption{Performance of the ablation study. Mean and standard deviation (std) of performance values in 20 runs. Setting 1: quality maps of source video are removed, Setting 2: quality maps of transcoded videos are removed.}
\label{tab:ablation_study}
\centering
\begin{tabular}{|c|c|c|c|}
\hline
Method  & SROCC              & PLCC               & RMSE               \\ \hline
Full version & 0.853 ($\pm$0.039) & 0.878 ($\pm$0.040) & 0.409 ($\pm$0.070) \\ \hline
Setting 1   & 0.806 ($\pm$0.058) & 0.835 ($\pm$0.057) & 0.478 ($\pm$0.073) \\ \hline
Setting 2   & 0.810 ($\pm$0.043) & 0.839 ($\pm$0.049) & 0.478 ($\pm$0.074) \\ \hline
\end{tabular}
\end{table}

\begin{figure}[t]
\centering
\includegraphics[width=1.0\columnwidth]{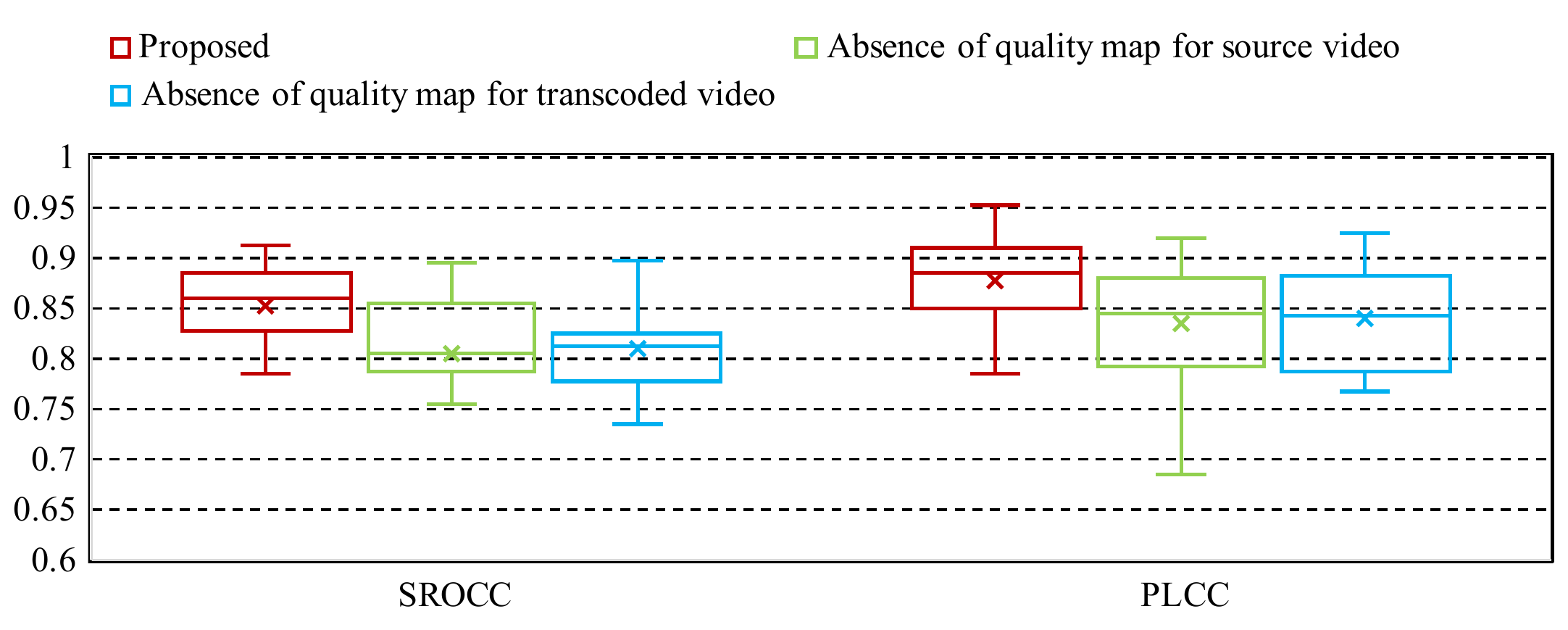}
\caption{Box plot in the ablation studies. The marks $\times$ in the middle represents the average. The bottom, middle and top bounds of the box represent the 25\%, 50\% and 75\% percentage points, respectively.}
\label{Fig:ablation_box_plot}
\end{figure}

To further provide evidence regarding the effectiveness of the proposed framework, we have conducted ablation studies by removing the quality maps of source and transcoded videos.

\subsubsection{Absence of quality map for source video}
\label{sssec:ablation_sourcemap}

We first show the performance variation due to the removal of the quality maps of source videos. In particular, these quality maps are replaced by source video frames, such that frames of source videos and quality maps of transcoded videos are fed to the pooling network.

\subsubsection{Absence of quality map for transcoded video}
\label{sssec:ablation_transmap}

The performance variations due to the removal of the quality maps of transcoded videos are further studied.  In this manner, the quality maps of source videos and frames of transcoded video are fed to the pooling network.

We compare the full version of our proposed method (red) with source video quality map removed configuration (green) and transcoded video quality map removed configuration (blue), as shown in Table~\ref{tab:ablation_study} and Fig.~\ref{Fig:ablation_box_plot}.
The removal of the source video quality maps or transcoded video quality maps causes significant performance drop, further verifying the effectiveness of quality maps of source videos or transcoded videos. 

\section{Conclusions}
\label{sec:conclusion}

In this paper, we have systematically studied the video quality of UGC content. To facilitate the development of VQA for UGC videos, we have constructed a new subjective quality database. This database contains diverse UGC video sources along with their transcoded versions under different compression standards and levels. The subjective ratings of these videos are also provided as the ground truth. Based on the interesting observations from the developed database, we propose a new objective video quality model with the design philosophy that the quality prediction does not only rely on the divergence of source video and transcoded video, but also the intrinsic quality of the source videos. 
The experimental results show that our method outperforms the state-of-the-art quality assessment methods. The proposed VQA method is also envisioned to be further adopted to regularize the quality of the output UGC videos, in an effort to provide a new paradigm of quality driven UGC video coding. 


\ifCLASSOPTIONcaptionsoff
  \newpage
\fi



\bibliographystyle{IEEEtran}
\bibliography{refs}
%



%




\end{document}